%% file: main-split-join.tex
\documentclass[envcountsame]{llncs}%

\usepackage{times}
\usepackage{amsmath}
\usepackage{amssymb}
\usepackage{tikz}
\usepackage{stmaryrd}
\usepackage{btran}
\usepackage{tikz}
\usepackage{psfrag}
\usepackage{ifthen}
\newcommand{\techRep}{true} %% switch here between true and false
\newcommand{\iftechrep}{\ifthenelse{\equal{\techRep}{true}}}

\newcommand{\PTime}{\textup{P}}
\newcommand{\PSPACE}{\textup{PSPACE}}
\newcommand{\EXPTIME}{\mbox{\textup{EXPTIME}}}
\newcommand{\PosSLP}{\textup{PosSLP}}
\newcommand{\tran}[1]{\xrightarrow{#1}}
\newcommand{\Bas}{{\cal B}}
\newcommand{\N}{\mathbb{N}}
\newcommand{\Q}{\mathbb{Q}}
\newcommand{\R}{\mathbb{R}}
\newcommand{\ExThR}{\mathit{ExTh}(\R)}
\newcommand{\calF}{{\cal F}}
\newcommand{\calP}{{\cal P}}
\renewcommand{\P}[1]{\calP\left(#1\right)}
\newcommand{\Ex}[1]{\mathbb{E}\left[#1\right]}
\newcommand{\E}{\mathbb{E}}
\newcommand{\Front}{\mathit{Front}}
\newcommand{\Prob}{\mathit{Prob}}
\newcommand{\Run}{\mathit{Run}}
\newcommand{\tRun}[2]{\Run(#1\mathord{\downarrow}#2)}
\newcommand{\tRunM}[3]{\Run[#1](#2\mathord{\downarrow}#3)}
\newcommand{\term}[2]{[#1\mathord{\downarrow}#2]}
\newcommand{\termV}[2]{\llbracket#1\mathord{\downarrow}#2\rrbracket}
\newcommand{\Tim}[1]{\mathbf{T}_{\!#1}}
\newcommand{\Wor}[1]{\mathbf{W}_{\!\!#1}}
\newcommand{\Spa}[1]{\mathbf{S}_{#1}}
\newcommand{\<}{\langle}
\renewcommand{\>}{\rangle}
\newcommand{\bs}[1]{\llparenthesis #1 \rrparenthesis}
\newcommand{\barS}{\overline{S}}
\newcommand{\barSigma}{\overline{\Sigma}}
\newcommand{\barGamma}{\overline{\Gamma}}
\newcommand{\barQ}{\overline{Q}}

\newcommand{\norm}[1]{\left\| #1 \right\|}
\newcommand{\es}[1]{\boldsymbol{e}^{(#1)}}
\newcommand{\qs}[1]{\boldsymbol{q}^{(#1)}}
\newcommand{\zs}[1]{\boldsymbol{z}^{(#1)}}
\newcommand{\rs}[1]{\boldsymbol{r}^{(#1)}}
\newcommand{\Ls}[1]{L^{(#1)}}%
\newcommand{\vf}{\boldsymbol{f}}%
\newcommand{\vu}{\boldsymbol{u}}%
\newcommand{\vx}{\boldsymbol{x}}%
\newcommand{\vy}{\boldsymbol{y}}%
\newcommand{\vz}{\boldsymbol{z}}%
\newcommand{\vw}{\boldsymbol{w}}%
\newcommand{\vzero}{\boldsymbol{0}}%
\newcommand{\vone}{\boldsymbol{1}}%

\newenvironment{qtheorem}[1]{%
{\mbox{}\newline\noindent\bf Theorem #1.}
\begin{itshape}%
}{%
\end{itshape}%
}
\newenvironment{qlemma}[1]{%
{\par\mbox{}\newline\noindent\bf Lemma #1.}%
\begin{itshape}%
}{%
\end{itshape}%
}
\newenvironment{qcorollary}[1]{%
{\mbox{}\newline\noindent\bf Corollary #1.}
\begin{itshape}%
}{%
\end{itshape}%
}

\newenvironment{qproposition}[1]{%
{\mbox{}\newline\noindent\bf Proposition #1.}
\begin{itshape}%
}{%
\end{itshape}%
}

\title{On Probabilistic Parallel Programs with \\ Process Creation and Synchronisation
\thanks{The first author is supported by a postdoctoral fellowship of the German Academic Exchange Service (DAAD).
The second author is supported by EPSRC grant EP/G050112/1.}}
\author{Stefan Kiefer \and Dominik Wojtczak}
\institute{Oxford University Computing Laboratory, UK
\email{\{stefan.kiefer,dominik.wojtczak\}@comlab.ox.ac.uk}}

\begin{document}
%---------------------
\sloppy
\maketitle

\begin{abstract}
 We initiate the study of probabilistic parallel programs with dynamic process creation and synchronisation.
 To this end, we introduce \emph{probabilistic split-join systems (pSJSs)}, a model for parallel programs,
  generalising both probabilistic pushdown systems (a model for sequential probabilistic procedural programs which is equivalent to recursive Markov chains)
  and stochastic branching processes (a classical mathematical model with applications in various areas such as biology, physics, and language processing).
 Our pSJS model allows for a possibly recursive spawning of parallel processes;
  the spawned processes can synchronise and return values.
 We study the basic performance measures of pSJSs, especially the distribution and expectation of space, work and time.
 Our results extend and improve previously known results on the subsumed models.
 We also show how to do performance analysis in practice,
  and present two case studies illustrating the modelling power of pSJSs.
\end{abstract}

\input{intro}
\input{prelim}
\input{theoretical}
\input{practical}
\input{conclusions}

\medskip
\noindent \emph{Acknowledgements.} We thank Javier Esparza, Alastair Donaldson, Markus M\"uller-Olm, Luke Ong and Thomas Wahl
 for helpful discussions on the non-probabilistic version of pSJSs.
We also thank the anonymous referees for valuable comments.
%---------------------
\bibliographystyle{plain} %oder alpha oder splncs
\bibliography{db}

\iftechrep{
\newpage
\appendix

\input{app-pPDS}
\input{app-termination}
\input{app-finite-space}
\input{app-work-time}
}

\end{document}

%% file: intro.tex
\section{Introduction} \label{sec:intro}

The verification of probabilistic programs with possibly recursive procedures has been intensely studied in the last years.
The Markov chains or Markov Decision Processes underlying these systems may have infinitely many states.
Despite this fact, which prevents the direct application of the rich theory of finite Markov chains, many positive results have been obtained.
Model-checking algorithms have been proposed for both linear and branching temporal logics \cite{EKM04,EYstacs05Extended,YannakakisE05},
 algorithms deciding properties of several kinds of games have been described (see e.g.~\cite{EtessamiY08}),
 and distributions and expectations of performance measures such as run-time and memory consumption have been investigated~\cite{EKM05,BEK09:fsttcs,BKKV10}.

In all these papers programs are modelled as \emph{probabilistic pushdown systems (pPDSs)} or, equivalently~\cite{EsparzaE04}, as recursive Markov chains.
Loosely speaking, a pPDS is a pushdown automaton whose transitions carry probabilities.
The \emph{configurations} of a pPDS are pairs containing the current control state and the current stack content.
In each \emph{step}, a new configuration is obtained from its predecessor by applying a transition rule,
 which may modify the control state and the top of the stack.

The programs modelled by pPDSs are necessarily sequential:
 at each point in time, only the procedure represented by the topmost stack symbol is active.
Recursion, however, is a useful language feature also for multithreaded and other parallel programming languages,
 such as Cilk and JCilk, which allow, e.g., for a natural parallelisation of divide-and-conquer algorithms~\cite{YBWCilk02,DanaherLeLe06}.
To model parallel programs in probabilistic scenarios, one may be tempted to use \emph{stochastic multitype branching processes},
 a classical mathematical model with applications in numerous fields including biology, physics and natural language processing~\cite{Harris63,AthreyaNey72}.
In this model, each process has a type, and each type is associated with a probability distribution on transition rules.
% describing whether a process of that type spawns other processes, changes its type, or terminates.
For instance, a branching process with the transition rules $X \btran{2/3} \{\}$, $X \btran{1/3} \{X,Y\}$, $Y \btran{1} \{X\}$
 can be thought of describing a parallel program with two types of processes, $X$ and $Y$.
A process of type~$X$ terminates with probability~$2/3$, and with probability~$1/3$ stays active and spawns a new process of type~$Y$.
A process of type~$Y$ changes its type to~$X$.
A \emph{configuration} of a branching process consists of a pool of currently active processes.
In each \emph{step}, all active processes develop in parallel, each one according to a rule which is chosen probabilistically.
For instance, a step transforms the configuration $\<XY\>$ into $\<XYX\>$ with probability $\frac13 \cdot 1$,
 by applying the second $X$-rule to the $X$-process and, in parallel, the $Y$-rule to the $Y$-process.

Branching processes do not satisfactorily model parallel programs,
 because they lack two key features: synchronisation and returning values.
In this paper we introduce \emph{probabilistic split-join systems (pSJSs)}, a model which offers these features.
%This is achieved by extending branching processes:
Parallel spawns are modelled by rules of the form $X \btran{} \<Y Z\>$.
The spawned processes $Y$ and~$Z$ develop independently;
 e.g., a rule $Y \btran{} Y'$ may be applied to the $Y$-process, replacing $Y$ by~$Y'$.
When terminating, a process enters a \emph{synchronisation state}, e.g.\ with rules $Y' \btran{} q$ and $Z \btran{} r$ (where $q$ and~$r$ are synchronisation states).
Once a process terminates in a synchronisation state, it waits for its \emph{sibling} to terminate in a synchronisation state as well.
In the above example, the spawned processes wait for each other, until they terminate in $q$ and~$r$.
At that point, they may \emph{join} to form a single process, e.g.\ with a rule $\<q r\> \btran{} W$.
So, synchronisation is achieved by the siblings waiting for each other to terminate.
All rules could be probabilistic.
Notice that synchronisation states can be used to return values;
 e.g., if the $Y$-process returns $q'$ instead of~$q$, this can be recorded by the existence of a rule $\<q' r\> \btran{} W'$,
  so that the resulting process (i.e., $W$ or~$W'$) depends on the values computed by the joined processes.
For the notion of siblings to make sense, a \emph{configuration} of a pSJS is not a set, but a binary tree whose leaves are \emph{process symbols}
 (such as $X,Y,Z$) or synchronisation states (such as $q,r$).
A \emph{step} transforms the leaves of the binary tree in parallel by applying rules;
 if a leaf is not a process symbol but a synchronisation state, it remains unchanged unless its sibling is also a synchronisation state
  and a joining rule (such as $\<q r\> \btran{} W$) exists, which removes the siblings and replaces their parent node with the right hand side.

\medskip
\noindent \emph{Related work.}
The probabilistic models closest to ours are pPDSs, recursive Markov chains, and stochastic branching processes, as described above.
The non-probabilistic (i.e., nondeterministic) version of pSJSs (SJSs, say) can be regarded as a special case of \emph{ground tree rewriting systems},
 see \cite{Loeding06} and the references therein.
A configuration of a ground tree rewriting system is a node-labelled tree, and a rewrite rule replaces a subtree.
The \emph{process rewrite system} (PRS) hierarchy of~\cite{Mayr00} features sequential and parallel process composition.
Due to its syntactic differences, it is not obvious whether SJSs are in that hierarchy.
They would be above pushdown systems (which is the sequential fragment of PRSs), because SJSs subsume pushdown systems,
 as we show in Section~\ref{sub:rel-pPDS} for the probabilistic models.
\emph{Dynamic pushdown networks} (DPNs)~\cite{BMT05} are a parallel extension of pushdown systems.
A configuration of a DPN is a list of configurations of pushdown systems running in parallel.
DPNs feature the spawning of parallel threads, and an extension of DPNs, called \emph{constrained DPNs},
 can also model joins via regular expressions on spawned children.
The DPN model is more powerful and more complicated than SJSs.
All those models are non-probabilistic.

\medskip
\noindent \emph{Organisation of the paper.}
In Section~\ref{sec:prelim} we formally define our model and provide further preliminaries.
Section~\ref{sec:theoretical} contains our main results:
 we study the relationship between pSJSs and pPDSs (Section~\ref{sub:rel-pPDS}),
 we show how to compute the probabilities for termination and finite space, respectively (Sections \ref{sub:termination-probabilities} and~\ref{sub:finite-space}),
 and investigate the distribution and expectation of work and time (Section~\ref{sub:work-time}).
In Section~\ref{sec:practical} we present two case studies illustrating the modelling power of pSJSs.
We conclude in Section 5.
All proofs are provided in \iftechrep{the appendix}{a technical report~\cite{KW11:tacas-report}}.

%% file: prelim.tex
\section{Preliminaries} \label{sec:prelim}

%Let $\N$ denote the set of nonnegative integers.
For a finite or infinite word~$w$, we write $w(0), w(1), \ldots$ to refer to its individual letters.
We assume throughout the paper that $\Bas$ is a fixed infinite set of \emph{basic process symbols}.
We use the symbols `$\<$' and `$\>$' as special letters not contained in~$\Bas$.
For an alphabet~$\Sigma$,
 we write $\<\Sigma \Sigma\>$ to denote the language $\{\<\sigma_1 \sigma_2\> \mid \sigma_1, \sigma_2 \in \Sigma\}$
 and $\Sigma^{1,2}$ to denote $\Sigma \cup \<\Sigma \Sigma\>$.
To a set~$\Sigma$ we associate a set~$T(\Sigma)$ of binary trees whose leaves are labelled with elements of~$\Sigma$.
Formally, $T(\Sigma)$ is the smallest language that contains $\Sigma$ and $\<T(\Sigma) T(\Sigma)\>$.
 %satisfies $\<t_1 t_2\> \in T(\Sigma)$ for all $t_1, t_2 \in T(\Sigma)$.
For instance, $\<\<\sigma \sigma\>\sigma\> \in T(\{\sigma\})$.

\begin{definition}[pSJS]
 Let $Q$ be a finite set of \emph{synchronisation states} disjoint from~$\Bas$ and not containing `$\<$' or `$\>$'.
 Let $\Gamma$ be a finite set of \emph{process symbols}, such that $\Gamma \subset \Bas \cup \<Q Q\>$.
 Define the \emph{alphabet} $\Sigma := \Gamma \cup Q$.
 Let $\delta \subseteq \Gamma \times \Sigma^{1,2}$ be a transition relation.
 Let $\Prob : \delta \to (0,1]$ be a function so that for all $a \in \Gamma$ we have $\sum_{a \btran{} \alpha \in \delta} \Prob(a \btran{} \alpha) = 1$.
 Then the tuple $S  = (\Gamma, Q, \delta, \Prob)$ is a \emph{probabilistic split-join system} (pSJS).
% If $|Q| = 1$, then $S$ is said to be \emph{communication-less}.
 A %communication-less
  pSJS with $\Gamma \cap \<Q Q\> = \emptyset$ is called \emph{branching process}.
\end{definition}
We usually write $a \btran{p} \alpha$ instead of $\Prob(a \btran{} \alpha) = p$.
For technical reasons we allow branching processes of ``degree 3'', i.e.,
 branching processes where $\Sigma^{1,2}$ may be extended to
 $\Sigma^{1,2,3} := \Sigma^{1,2} \cup \{\<\sigma_1 \sigma_2 \sigma_3\> \mid \sigma_1, \sigma_2, \sigma_3 \in \Sigma\}$.
In branching processes, it is usually sufficient to have $|Q| = 1$.

A \emph{Markov chain} is a stochastic process that can be described by a triple \mbox{$M = (D,\mathord{\tran{}},\Prob)$}
where $D$ is a finite or countably infinite set of \emph{states},
$\mathord{\tran{}} \subseteq D \times D$ is a \emph{transition relation},
and $\Prob$ is a function which to each transition $s \tran{} t$ of~$M$
assigns its probability $\Prob(s \tran{} t) > 0$ so that
for every $s \in D$ we have $\sum_{s \tran{} t} \Prob(s \tran{} t) =
1$ (as usual, we write $s \tran{x} t$ instead of $\Prob(s \tran{} t) = x$).
A \emph{path} (or \emph{run}) in $M$ is a finite (or infinite, resp.) word $u \in D^+ \cup D^\omega$,
such that $u(i{-}1) \tran{} u(i)$ for every \mbox{$1 \leq i < |u|$}.
%We denote by $\Run[M]$ the set of all runs in~$M$.
The set of all runs that start with a given path~$u$ is denoted by $\Run[M](u)$ (or $\Run(u)$, if $M$ is understood).
%When $M$ is understood, we write $\Run$ (or $\Run(u)$) instead of $\Run[M]$ (or $\Run[M](u)$, resp.).
%
To every $s \in D$ we associate the probability
space $(\Run(s),\calF,\calP)$ where
$\calF$ is the \mbox{$\sigma$-field} generated by all \emph{basic cylinders}
$\Run(u)$ where $u$ is a path starting with~$s$, and
$\calP: \calF \rightarrow [0,1]$ is the unique probability measure such that
$\calP(\Run(u)) = \Pi_{i{=}1}^{|u|-1} x_i$ where
$u(i{-}1) \tran{x_i} u(i)$ for every $1 \leq i < |u|$.
%If $|u| = 1$, we put $\calP(\Run(u)) = 1$.
Only certain subsets of $\Run(s)$ are $\calP$-measurable, but in this paper we only
deal with ``safe'' subsets that are guaranteed to be in $\calF$.
If $\mathbf{X}_s$ is a random variable over $\Run(s)$, we write $\Ex{\mathbf{X}_s}$ for its expectation.
%Given $s\in D$ and $A\subseteq D$, we say \emph{$A$ is reachable from~$s$}, if
%$\calP(\{w\in \Run(s)\mid \exists i\geq 0: w(i)\in A\}) > 0$.
For $s, t \in D$, we define $\tRun{s}{t} := \{w \in \Run(s) \mid \exists i \ge 0 : w(i) = t\}$
 and $\term{s}{t} := \P{\tRun{s}{t}}$.

To a pSJS $S  = (\Gamma, Q, \delta, \Prob)$ with alphabet $\Sigma = \Gamma \cup Q$
 we associate a Markov chain $M_S$ with $T(\Sigma)$ as set of states.
For $t \in T(\Sigma)$, we define $\Front(t) = a_1, \ldots, a_k$ as the unique finite sequence of subwords of~$t$ (read from left to right)
 with $a_i \in \Gamma$ for all $1 \le i \le k$.
We write $|\Front(t)| = k$.
If $k = 0$, then $t$ is called \emph{terminal}.
The Markov chain $M_S$ has a transition $t \tran{p} t'$, if:
 $\Front(t) = a_1, \ldots, a_k$;
 $a_i \btran{p_i} \alpha_i$ are transitions in~$S$ for all~$i$;
 $t'$ is obtained from~$t$ by replacing $a_i$ with $\alpha_i$ for all~$i$;
 and $p = \prod_{i=1}^k p_i$.
Note that $t \tran{1} t$, if $t$ is terminal.
For branching processes of degree~3, the set~$T(\Sigma)$ is extended in the obvious way
 to trees whose nodes may have two or three children.

%For $\sigma \in \Sigma$ and $q \in Q$ we define $\tRun{\sigma}{q} := \{w \in \Run(\sigma) \mid \exists i \ge 0 : w(i) = q\}$
% and $\term{\sigma}{q} := \P{\tRun{\sigma}{q}}$.
Denote by $\Tim{\sigma}$ a random variable over~$\Run(\sigma)$ where $\Tim{\sigma}(w)$ is either the least~$i \in \N$ such that $w(i)$ is terminal,
 or~$\infty$, if no such~$i$ exists.
Intuitively, $\Tim{\sigma}(w)$ is the number of steps in which $w$ terminates, i.e., the \emph{termination time}.
Denote by $\Wor{\sigma}$ a random variable over~$\Run(\sigma)$ where $\Wor{\sigma}(w) := \sum_{i=0}^\infty |\Front(w(i))|$.
Intuitively, $\Wor{\sigma}(w)$ is the total \emph{work} in~$w$.
Denote by~$\Spa{\sigma}$ a random variable over~$\Run(\sigma)$ where $\Spa{\sigma}(w) := \sup_{i=0}^\infty |w(i)|$,
 and $|w(i)|$ is the length of~$w(i)$ not counting the symbols `$\<$' and `$\>$'.
Intuitively, $\Spa{\sigma}(w)$ is the maximal number of processes during the computation, or, short, the \emph{space} of~$w$.

\begin{figure}
\iftechrep{}{\vspace{-10mm}}
  \begin{center}
   \begin{tikzpicture}
    \node (0)  at (0,0) {$X$};
    \node (1a) at (-1,-0.8) {$X$};
    \node (1b) at (+1,-0.8) {$X$};
    \node (2a) at (-1,-1.6) {$q$};
    \node (2b) at (+1,-1.6) {$r$};
    \node (3)  at (0,-2.4) {$X$};
    \node (4)  at (0,-3.2) {$q$};
    \draw (0)  -- (1a);
    \draw (0)  -- (1b);
    \draw (1a) -- (2a);
    \draw (1b) -- (2b);
    \draw (2a) -- (3);
    \draw (2b) -- (3);
    \draw (3) --  (4);
    \node (A)  at (5,0) {$X$};
    \node (Ba) at (4,-0.8) {$X$};
    \node (Bb) at (6,-0.8) {$X$};
    \node (Ca) at (4,-1.6) {$q$};
    \node (Cb) at (5.5,-1.6) {$X$};
    \node (Cc) at (6.5,-1.6) {$X$};
    \node (Db) at (5.5,-2.4) {$r$};
    \node (Dc) at (6.5,-2.4) {$q$};
    \draw (A) -- (Ba);
    \draw (A) -- (Bb);
    \draw (Ba) -- (Ca);
    \draw (Bb) -- (Cb);
    \draw (Bb) -- (Cc);
    \draw (Cb) -- (Db);
    \draw (Cc) -- (Dc);
   \end{tikzpicture}
  \end{center}
\iftechrep{}{\vspace{-8mm}}
  \caption{Two terminating runs}
  \label{fig:ex-run}
\end{figure}
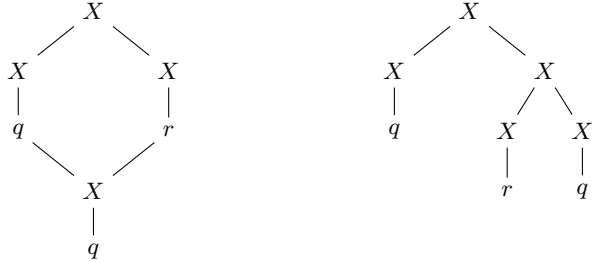

\begin{example}
 Consider the pSJS with $\Gamma = \{X, \<q r\>\}$ and $Q = \{q,r\}$ and the transitions
  $X \btran{0.5} \<X X\>$, \ $X \btran{0.3} q$, \ $X \btran{0.2} r$, \ $\<q r\> \btran{1} X$.
 Let $u = X \ \<X X\> \ \<q r\> \ X \ q \ q$.
 Then $u$ is a path, because we have $X \tran{0.5} \<X X\> \tran{0.06} \<q r\> \tran{1} X \tran{0.3} q \tran{1} q$.
 Note that $q$ is terminal.
 The set $\Run(u)$ contains only one run, namely $w := u(0) u(1) u(2) u(3) u(4) u(4) \cdots$.
 We have $\P{\Run(u)} = 0.5 \cdot 0.06 \cdot 0.3$, and $\Tim{X}(w) = 4$, $\Wor{X}(w) = 5$, and $\Spa{X}(w) = 2$.
 The dags in Figure~\ref{fig:ex-run} graphically represent this run (on the left),
  and another example run (on the right) with $\Tim{X} = 3$, $\Wor{X} = 5$, and $\Spa{X} = 3$.
\end{example}

\begin{remark}
 Our definition of pSJSs may be more general than needed from a modelling perspective:
  e.g., our rules allow for both synchronisation and splitting in a single step.
 We choose this definition for technical convenience and to allow for easy comparisons with pPDSs (Section~\ref{sub:rel-pPDS}).
\end{remark}

The complexity-theoretic statements in this paper are with respect to the \emph{size} of the given pSJS $S = (\Gamma, Q, \delta, \Prob)$,
 which is defined as $|\Gamma| + |Q| + |\delta| + |\Prob|$,
 where $|\Prob|$ equals the sum of the sizes of the binary representations of the values of~$\Prob$.
A formula of~$\ExThR$, the existential fragment of the first-order theory of the reals, is of the form
 $\exists x_1 \ldots \exists x_m R(x_1, \ldots, x_n)$, where $R(x_1, \ldots, x_n)$ is a boolean combination of comparisons of the form
 $p(x_1, \ldots, x_n) \sim 0$, where $p(x_1, \ldots, x_n)$ is a multivariate polynomial and
  $\mathord{\sim} \in \{ \mathord{<}, \mathord{>}, \mathord{\le}, \mathord{\ge}, \mathord{=}, \mathord{\ne} \}$.
The validity of closed formulas ($m=n$) is decidable in \PSPACE~\cite{Can88,Renegar92}.
We say that one can \emph{efficiently express} a value $c \in \R$ associated with a pSJS,
 if one can, in polynomial space, construct a formula $\phi(x)$ in~$\ExThR$ of polynomial length such that $x$ is the only free variable in~$\phi(x)$,
 and $\phi(x)$ is true if and only if $x=c$.
Notice that if $c$ is efficiently expressible, then $c \sim \tau$ for $\tau \in \Q$ is decidable in \PSPACE\ for
 $\mathord{\sim} \in \{ \mathord{<}, \mathord{>}, \mathord{\le}, \mathord{\ge}, \mathord{=}, \mathord{\ne} \}$.

For some lower bounds, we prove hardness (with respect to P-time many-one reductions) in terms of the \PosSLP\ decision problem.
The \PosSLP\ (Positive Straight-Line Program) problem asks whether a given straight-line program or,
 equivalently, arithmetic circuit with operations $\mathord{+}$, $\mathord{-}$, $\mathord{\cdot}$, and inputs 0 and 1, and a designated output gate,
 outputs a positive integer or not.
\PosSLP\ is in \PSPACE.
More precisely, it is known to be on the 4th level of the Counting Hierarchy~\cite{AllenderBKM06};
 it is not known to be in~NP.
The \PosSLP\ problem is a fundamental problem for numerical computation;
 it is complete for the class of decision problems that can be solved in polynomial time on models
 with unit-cost exact rational arithmetic, see \cite{AllenderBKM06,EYstacs05Extended} for more details.

%% file: theoretical.tex
\section{Results} \label{sec:theoretical}

\subsection{Relationship with probabilistic pushdown systems (pPDSs)} \label{sub:rel-pPDS}

%Probabilistic pushdown systems (pPDSs) and, equivalently~\cite{EsparzaE04}, recursive Markov chains
% both model probabilistic sequential programs with possibly recursive procedure calls
% and are very well studied \cite{EKM04,YannakakisE05,BrazdilEK05,EKM05,EtessamiY08,EYstacs05Extended,BEK09:fsttcs}.
We show that pSJSs subsume pPDSs.
A \emph{probabilistic pushdown system (pPDS)}
\cite{EKM04,EKM05,BEK09:fsttcs,BKKV10}
is a tuple $S = (\Gamma,Q,\delta,\Prob)$,
 where $\Gamma$ is a finite \emph{stack alphabet}, $Q$ is a finite set of \emph{control states},
 $\delta \subseteq Q \times \Gamma \times Q \times \Gamma^{\leq 2}$
  (where $\Gamma^{\leq 2}=\{\alpha\in \Gamma^*, |\alpha|\leq 2\}$)
  is a \emph{transition relation},
 and $\Prob: \delta \to (0,1]$ is a function so that for all $q \in Q$ and $a \in \Gamma$
  we have $\sum_{q a \btran{} r \alpha} \Prob(q a \btran{} r \alpha) = 1$.
One usually writes $q a \tran{p} r \alpha$ instead of $\Prob(q a \tran{} r \alpha) = p$.
To a pPDS $S = (\Gamma, Q, \delta, \Prob)$
 one associates a Markov chain $M_S$ with $Q \times \Gamma^*$ as set of states,
 and transitions $q \tran{1} q $ for all $q \in Q$,
 and $q a \beta \tran{p} r \alpha \beta$ for all $q a \btran{p} r \alpha$ and all $\beta \in \Gamma^*$.

A pPDS~$S_P$ with $\Gamma_P$ as stack alphabet, $Q_P$ as set of control states, and transitions~$\mathord{\btran{p}_P}$
 can be transformed to an equivalent pSJS~$S$:
Take $Q := Q_P \cup \Gamma_P$ as synchronisation states;
 $\Gamma := \{\< q a \> \mid q \in Q_P, \ a \in \Gamma_P\}$ as process symbols;
 and transitions $\< q a \> \btran{p} \< \< r b \> c \>$ for all $q a \btran{p}_P r b c$,
  $\< q a \> \btran{p} \< r b \>$ for all $q a \btran{p}_P r b$, and
  $\< q a \> \btran{p} r$ for all $q a \btran{p}_P r$.
The Markov chains $M_{S_P}$ and $M_S$ are isomorphic.
Therefore, we occasionally say that a pSJS \emph{is} a pPDS, if it can be obtained from a pPDS by this transformation.
Observe that in pPDSs, we have $\Tim{} = \Wor{}$, because there is no parallelism.

Conversely, a pSJS~$S$ with alphabet $\Sigma = \Gamma \cup Q$
 can be transformed into a pPDS~$S_P$ by ``serialising''~$S$:
Take $Q_P := \{ \Box \} \cup \{\overline{q} \mid q \in Q\}$ as control states;
 $\Gamma_P := \Gamma \cup Q \cup \{\widetilde{q} \mid q \in Q\}$ as stack alphabet;
 and transitions $\Box a \btran{p}_P \Box \sigma_1 \sigma_2$ for all $a \btran{p} \<\sigma_1 \sigma_2\>$,
 $\Box a \btran{p}_P \Box \sigma$ for all $a \btran{p} \sigma$ with $\sigma \in \Sigma \setminus \<Q Q\>$,
 and $\Box q \btran{1}_P \overline{q}$ for all $q \in Q$,
 and $\overline{q} \sigma \btran{1}_P \Box \sigma \widetilde{q}$ for all $q \in Q$ and $\sigma \in \Sigma$,
 and $\overline{r} \widetilde{q} \btran{1}_P \Box \<q r\>$ for all $q, r \in Q$.
The Markov chains $M_S$ and $M_{S_P}$ are \emph{not} isomorphic.
However, we have:
\newcommand{\stmtpropserialisation}{
 There is a probability-preserving bijection
  between the runs $\tRun{\sigma}{q}$ in~$M_S$ and the runs $\tRun{\Box \sigma}{\overline{q}}$ in~$M_{S_P}$.
 In particular, we have $\term{\sigma}{q} = \term{\Box \sigma}{\overline{q}}$.
}
\begin{proposition} \label{prop:serialisation}
 \stmtpropserialisation
\end{proposition}
For example, the pSJS run on the left side of Figure~\ref{fig:ex-run} corresponds to the pPDS run
 $\Box X \tran{0.5} \Box X X \tran{0.3} \Box q X \tran{1} \overline{q} X \tran{1} \Box X \widetilde{q} \tran{0.2} \Box r \widetilde{q}
  \tran{1} \overline{r} \widetilde{q} \tran{1} \Box \<q r\> \tran{1} \Box X \tran{0.3} \Box q \tran{1} 
  \overline{q} \tran{1} \overline{q} \tran{1} \ldots$

\subsection{Probability of Termination} \label{sub:termination-probabilities}

We call a run \emph{terminating}, if it reaches a terminal tree.
Such a tree can be a single synchronisation state (e.g., $q$ on the left of Figure~\ref{fig:ex-run}),
 or another terminal tree (e.g., $\<q \<r q\> \>$ on the right of Figure~\ref{fig:ex-run}).
For any $\sigma \in \Sigma$, we denote by $\term{\sigma}{}$ the termination probability when starting in~$\sigma$;
 i.e., $\term{\sigma}{} = \sum_{t \text{ is terminal}} \term{\sigma}{t}$.
One can transform any pSJS~$S$ into a pSJS~$S'$ such that whenever a run in~$S$ terminates,
 then a corresponding run in~$S'$ terminates in a synchronisation state.
This transformation is by adding a fresh state $\check{q}$, and transitions $\<rs\> \btran{1} \check{q}$ for all $r,s \in Q$ with $\<rs\> \not\in \Gamma$,
 and $\<\check{q} r\> \btran{1} \check{q}$ and $\<r \check{q}\> \btran{1} \check{q}$ for all $r \in Q$.
It is easy to see that this keeps the probability of termination unchanged, and modifies the random variables $\Tim{\sigma}$ and~$\Wor{\sigma}$
 by at most a factor~$2$.
Notice that the transformation can be performed in polynomial time.
After the transformation we have $\term{\sigma}{} = \sum_{q \in Q} \term{\sigma}{q}$.
A pSJS which satisfies this equality will be called \emph{normalised} in the following.
%Therefore, we will focus on runs in~$\tRun{\sigma}{q}$ in the following.
From a modelling point of view, pSJSs may be expected to be normalised in the first place:
 a terminating program should terminate all its processes.
% it may be bad, if the created processes still exist after the program has terminated.

We set up an equation system for the probabilities $\term{\sigma}{q}$.
For each $\sigma \in \Sigma$ and $q \in Q$, the equation system has a variable of the form $\termV{\sigma}{q}$ and an equation of the form
 $\termV{\sigma}{q} = f_{\termV{\sigma}{q}}$, where $f_{\termV{\sigma}{q}}$ is a multivariate polynomial with nonnegative coefficients.
More concretely:
%Set up the following equation system with variables of the form $\termV{\sigma}{q}$ where $\sigma \in \Sigma$ and $q \in Q$.
If $q \in Q$, then we set $\termV{q}{q} = 1$;
if $r \in Q \setminus \{q\}$, then we set $\termV{r}{q} = 0$;
if $a \in \Gamma$, then we set
\begin{align*}
% \termV{q}{q} & = 1 \qquad \text{and } \qquad \termV{r}{q} = 0 \quad \text{if $r \in Q \setminus \{q\}$} \\
 \termV{a}{q} & = \mathop{\sum_{a \btran{p} \< \sigma_1 \sigma_2 \>}}_{\<q_1 q_2\> \in \Gamma \cap \<Q Q\>}
                   p \cdot \termV{\sigma_1}{q_1} \cdot \termV{\sigma_2}{q_2} \cdot \termV{\<q_1 q_2\>}{q} \ +
                  \mathop{\sum_{a \btran{p} \sigma'}}_{\sigma' \in \Sigma \setminus \<Q Q\>} p \cdot \termV{\sigma'}{q} \,.
\end{align*}

\newcommand{\stmtproptermination}{
 Let $\sigma \in \Sigma$ and $q \in Q$.
 Then $\term{\sigma}{q}$ is the value for $\termV{\sigma}{q}$ in the least (w.r.t.\ componentwise ordering)
  nonnegative solution of the above equation system.
}
\begin{proposition} \label{prop:termination}
 \stmtproptermination
\end{proposition}
One can efficiently approximate~$\term{\sigma}{q}$ by applying Newton's method
 to the fixed-point equation system from Proposition~\ref{prop:termination}, cf.~\cite{EYstacs05Extended}.
The convergence speed of Newton's method for such equation systems was recently studied in detail~\cite{EKL10:SICOMP}.
The simpler ``Kleene'' method (sometimes called ``fixed-point iteration'') often suffices, but can be much slower.
In the case studies of Section~\ref{sec:practical}, using Kleene for computing the termination probabilities up to machine accuracy was not a bottleneck.
\newcommand{\stmtthmterminationprobabilities}{
 Consider a pSJS with alphabet $\Sigma = \Gamma \cup Q$.
 Let $\sigma \in \Sigma$ and $q \in Q$.
 Then (1) one can efficiently express {(in the sense defined in Section~\ref{sec:prelim}) the value of }{}$\term{\sigma}{q}$,
 (2) deciding whether $\term{\sigma}{q} = 0$ is in~\PTime,
 and (3) deciding whether $\term{\sigma}{q} < 1$ is \PosSLP-hard even for pPDSs.
}
The following theorem essentially follows from similar results for pPDSs:
\begin{theorem}[cf.~\cite{EtessamiY05,EYstacs05Extended}] \label{thm:termination-probabilities}
 \stmtthmterminationprobabilities
\end{theorem}

\subsection{Probability of Finite Space} \label{sub:finite-space}

A run~$w \in \Run(\sigma)$ is either (i) terminating, or (ii) nonterminating with $\Spa{\sigma} < \infty$,
 or (iii) nonterminating with $\Spa{\sigma} = \infty$.
From a modelling point of view, some programs may be considered incorrect, if they do not terminate with probability~$1$.
As is well-known, this does not apply to programs like operating systems, network servers, system daemons, etc.,
 where nontermination may be tolerated or desirable.
Such programs may be expected not to need an infinite amount of space;
 i.e., $\Spa{\sigma}$ should be finite.

Given a pSJS~$S$ with alphabet $\Sigma = \Gamma \cup Q$, we show how to construct, in polynomial time,
 a normalised pSJS~$\barS$ with alphabet $\barSigma = \barGamma \cup \barQ \supseteq \Sigma$ where $\barQ = Q \cup \{\overline{q}\}$
 for a fresh synchronisation state~$\overline{q}$, and $\P{\Spa{a} < \infty = \Tim{a} \mid \Run(a)} = \term{a}{\overline{q}}$
  for all $a \in \Gamma$.
Having done that, we can compute this probability according to Section~\ref{sub:termination-probabilities}.

For the construction, we can assume w.l.o.g.\ that $S$ has been normalised using the procedure of Section~\ref{sub:termination-probabilities}.
Let $U := \{a \in \Gamma \mid \forall n \in \N : \P{\Spa{a} > n} > 0\}$.
\newcommand{\stmtlemunboundedpolytime}{
 The set~$U$ can be computed in polynomial time.
}
\begin{lemma} \label{lem:unbounded-poly-time}
 \stmtlemunboundedpolytime
\end{lemma}
Let $B := \{a \in \Gamma \setminus U \mid \forall q \in Q : \term{a}{q} = 0\}$,
 so $B$ is the set of process symbols $a$ that are both ``bounded above'' (because $a \not\in U$) and ``bounded below'' (because $a$ cannot terminate).
By Theorem~\ref{thm:termination-probabilities}~(2) and Lemma~\ref{lem:unbounded-poly-time} we can compute~$B$ in polynomial time.
Now we construct~$\barS$ by modifying~$S$ as follows: we set $\barQ := Q \cup \{\overline{q}\}$ for a fresh synchronisation state~$\overline{q}$;
 we remove all transitions with symbols $b \in B$ on the left hand side and replace them with a new transition $b \btran{1} \overline{q}$;
 we add transitions $\<q_1 q_2\> \btran{1} \overline{q}$ for all $q_1, q_2 \in \barQ$ with $\overline{q} \in \{q_1, q_2\}$.
We have the following proposition.
\newcommand{\stmtpropfinitespace}{
(1) The pSJS~$\barS$ is normalised;
(2) the value $\term{a}{q}$ for $a \in \Gamma$ and $q \in Q$ is the same in $S$ and~$\barS$;
(3) we have $\P{\Spa{a} < \infty = \Tim{a} \mid \Run(a)} = \term{a}{\overline{q}}$ for all $a \in \Gamma$.
}
\begin{proposition} \label{prop:finite-space}
 \stmtpropfinitespace
\end{proposition}
Proposition~\ref{prop:finite-space} allows for the following theorem.
\newcommand{\stmtthmfinitespace}{
 Consider a pSJS with alphabet $\Sigma = \Gamma \cup Q$ and $a \in \Gamma$.
 Let $s := \P{\Spa{a} < \infty}$.
 Then (1) one can efficiently express~$s$,
 (2) deciding whether $s = 0$ is in~\PTime,
 and (3) deciding whether $s < 1$ is \PosSLP-hard even for pPDSs.
}
\begin{theorem} \label{thm:finite-space}
 \stmtthmfinitespace
\end{theorem}
Theorem~\ref{thm:finite-space}, applied to pPDSs, improves Corollary~6.3 of~\cite{EKM05}.
There it is shown for pPDSs that comparing $\P{\Spa{a} < \infty}$ with $\tau \in \Q$ is in~\EXPTIME, and in~\PSPACE\ if $\tau \in \{0,1\}$.
With Theorem~\ref{thm:finite-space} we get \PSPACE\ for $\tau \in \Q$, and \PTime\ for $\tau = 0$.

\subsection{Work and Time} \label{sub:work-time}

We show how to compute the distribution and expectation of work and time
 of a given pSJS~$S$ with alphabet $\Sigma = \Gamma \cup Q$.

\medskip
\noindent \emph{Distribution.}
For $\sigma \in \Sigma$ and $q \in Q$, let $T_{\sigma \downarrow q}(k) := \P{\tRun{\sigma}{q}, \ \Tim{\sigma}=k \mid \Run(\sigma)}$.
It is easy to see that, for $k \ge 1$ and $a \in \Gamma$ and $q \in Q$, we have
\begin{align*}
 T_{a \downarrow q}(k) & = \mathop{\sum_{a \btran{p} \< \sigma_1 \sigma_2 \>}}_{\<q_1 q_2\> \in \Gamma \cap \<Q Q\>}
                           p \cdot \mathop{\sum_{\ell_1, \ell_2, \ell_3 \ge 0}}_{\max\{\ell_1, \ell_2\} + \ell_3 = k-1}
                            T_{\sigma_1 \downarrow q_1}(\ell_1) \cdot T_{\sigma_2 \downarrow q_2}(\ell_2) \cdot T_{\< q_1 q_2 \> \downarrow q}(\ell_3) + \mbox{} \\
                       & \qquad \mathop{\sum_{a \btran{p} \sigma'}}_{\sigma' \in \Sigma \setminus \<Q Q\>} p \cdot T_{\sigma' \downarrow q}(k-1) \,.
\end{align*}
This allows to compute the distribution of time (and, similarly, work) using dynamic programming.
In particular, for any~$k$, one can compute
 $\overrightarrow{T}_{\sigma \downarrow q}(k) := \P{\Tim{\sigma} > k \mid \tRun{\sigma}{q}}
   = 1 -  \frac{1}{\term{\sigma}{q}} \sum_{i=0}^k T_{\sigma \downarrow q}(k)$.

\medskip
\noindent \emph{Expectation.}
For any random variable~$Z$ taking positive integers as value, it holds $\E Z = \sum_{k=0}^\infty \P{Z > k}$.
Hence, one can approximate $\Ex{\Tim{\sigma} \mid \tRun{\sigma}{q}} = \sum_{k=0}^\infty \overrightarrow{T}_{\sigma \downarrow q}(k)$
  by computing $\sum_{k=0}^\ell \overrightarrow{T}_{\sigma \downarrow q}(k)$ for large~$\ell$.
In the rest of the section we show how to decide on the finiteness of expected work and time.
It follows from Propositions \ref{prop:transformation} and~\ref{prop:bp-critical} below that the expected work $\Ex{\Wor{\sigma} \mid \tRun{\sigma}{q}}$
 is easier to compute: it is the solution of a linear equation system.

We construct a branching process~$\barS$ with process symbols $\barGamma = \{ \bs{a q} \mid a \in \Gamma, q \in Q, \term{a}{q} > 0 \}$,
 synchronisation states $\barQ = \{ \bot \}$
 and transitions as follows.
For notational convenience, we identify $\bot$ and $\bs{q q}$ for all $q \in Q$.
For $\bs{a q} \in \barGamma$, we set
\begin{itemize}
\item
 $\bs{a q} \btran{y/\term{a}{q}} \< \bs{\sigma_1 q_1} \bs{\sigma_2 q_2} \bs{\<q_1 q_2\> q} \>$
  for all $a \btran{p} \<\sigma_1 \sigma_2\>$ and $\< q_1 q_2 \> \in \Gamma \cap \<Q Q\>$, where
  $y := p \cdot \term{\sigma_1}{q_1} \cdot \term{\sigma_2}{q_2} \cdot \term{\< q_1 q_2 \>}{q} > 0$\,;
\item
 $\bs{a q} \btran{y/\term{a}{q}} \bs{\sigma' q}$
  for all $a \btran{p} \sigma'$ with $\sigma' \in \Sigma \setminus \<Q Q\>$, where
  $y := p \cdot \term{\sigma'}{q} > 0$\,.
\end{itemize}
The following proposition (inspired by a statement on pPDSs~\cite{BKKV10})
 links the distributions of $\Wor{\sigma}$ and~$\Tim{\sigma}$ conditioned under termination in~$q$
 with the distributions of $\Wor{\bs{\sigma q}}$ and~$\Tim{\bs{\sigma q}}$.
\newcommand{\stmtproptransformation}{%
 Let $\sigma \in \Sigma$ and $q \in Q$ with $\term{\sigma}{q} > 0$.
 Then
  \begin{align*}
   \P{\Wor{\sigma} = n \mid \tRun{\sigma}{q}}   & =   \P{\Wor{\bs{\sigma q}} =   n \mid \Run({\bs{\sigma q}})} && \text{for all $n \ge 0$} \quad \text{and} \\
   \P{\Tim{\sigma} \le n \mid \tRun{\sigma}{q}} & \le \P{\Tim{\bs{\sigma q}} \le n \mid \Run({\bs{\sigma q}})} && \text{for all $n \ge 0$.}
  \end{align*}
 In particular, we have $\term{\bs{\sigma q}}{} = 1$.
}
\begin{proposition} \label{prop:transformation}
 \stmtproptransformation
\end{proposition}
Proposition~\ref{prop:transformation} allows us to focus on branching processes.
For $X \in \Gamma$ and a finite sequence $\sigma_1, \ldots, \sigma_k$ with $\sigma_i \in \Sigma$,
 define $|\sigma_1, \ldots, \sigma_k|_X := |\{i \mid 1 \le i \le k, \ \sigma_i = X\}|$, i.e., the number of $X$-symbols in the sequence.
We define the \emph{characteristic matrix} $A \in \R^{\Gamma \times \Gamma}$ of a branching process by setting
  \[
   A_{X,Y} := \sum_{X \btran{p} \<\sigma_1 \sigma_2 \sigma_3\>} p \cdot |\sigma_1,\sigma_2,\sigma_3|_Y +
             \sum_{X \btran{p} \<\sigma_1 \sigma_2\>} p \cdot |\sigma_1,\sigma_2|_Y +
             \sum_{X \btran{p} \sigma_1} p \cdot |\sigma_1|_Y \,.
  \]
It is easy to see that %$A(X,Y) = \Ex{|\Front(w(1))|_Y \mid w \in \Run(X)}$; i.e.,
 the $(X,Y)$-entry of~$A$ is the expected number of $Y$-processes after the first step, if starting in a single~$X$-process.
If $S$ is a branching process and $X_0 \in \Gamma$, we call the pair $(S,X_0)$ a \emph{reduced branching process},
 if for all $X \in \Gamma$ there is $i \in \N$ such that $(A^i)_{X_0,X} > 0$.
Intuitively, $(S,X_0)$ is reduced, if, starting in~$X_0$, all process symbols can be reached with positive probability.
If $(S,X_0)$ is not reduced, it is easy to reduce it in polynomial time by eliminating all non-reachable process symbols.

The following proposition characterises the finiteness of both expected work and expected time in
 terms of the spectral radius $\rho(A)$ of~$A$.
(Recall that $\rho(A)$ is the largest absolute value of the eigenvalues of~$A$.)
\newcommand{\stmtpropbpcritical}{%
 Let $(S,X_0)$ be a reduced branching process. % with $\term{X_0}{} = 1$.
 Let $A$ be the associated characteristic matrix.
 Then the following statements are equivalent:
 \[ \text{(1)} \ \E \Wor{X_0} \text{ is finite;} \qquad \qquad \text{(2)} \ \E \Tim{X_0} \text{ is finite;} \qquad \qquad \text{(3)} \ \rho(A) < 1\,. \]
 Further, if $\E \Wor{X_0}$ is finite, then it equals the $X_0$-component of $(I-A)^{-1} \cdot \vone$,
  where $I$ is the identity matrix, and $\vone$ is the column vector with all ones.
}
\begin{proposition} \label{prop:bp-critical}
 \stmtpropbpcritical
\end{proposition}
Statements similar to Proposition~\ref{prop:bp-critical} do appear in the standard branching process literature \cite{Harris63,AthreyaNey72},
 however, not explicitly enough to cite directly or with stronger assumptions\footnote{For example, \cite{AthreyaNey72} assumes
 that there is~$n \in \N$ such that $A^n$ is positive in all entries,
 a restriction which is not natural for our setting.}.
Our proof adapts a technique which was developed in~\cite{BEK09:fsttcs} for a different purpose.
It uses only basic tools and Perron-Frobenius theory, the spectral theory of nonnegative matrices.
\newcommand{\stmtcorbpworktime}{
 Consider a branching process with process symbols~$\Gamma$ and $X_0 \in \Gamma$.
 Then $\E \Wor{X_0}$ and $\E \Tim{X_0}$ are both finite or both infinite.
 Distinguishing between those cases is in~\PTime.
}
Proposition~\ref{prop:bp-critical} has the following consequence:
\begin{corollary} \label{cor:bp-work-time}
 \stmtcorbpworktime
\end{corollary}
By combining the previous results we obtain the following theorem.
\newcommand{\stmtthmworktime}{%
 Consider a pSJS~$S$ with alphabet $\Sigma = \Gamma \cup Q$.
 Let $a \in \Gamma$.
 Then $\E \Wor{a}$ and $\E \Tim{a}$ are both finite or both infinite.
 Distinguishing between those cases is in~\PSPACE, and \PosSLP-hard even for pPDSs.
 Further, if $S$ is normalised and $\E \Wor{a}$ is finite, one can efficiently express $\E \Wor{a}$.
}
\begin{theorem} \label{thm:work-time}
 \stmtthmworktime
\end{theorem}
Theorem~\ref{thm:work-time} can be interpreted as saying that,
 although the pSJS model does not impose a bound on the number of active processes at a time,
 its parallelism \emph{cannot} be used to do an infinite expected amount of work in a finite expected time.
However, the ``speedup'' $\Ex{\Wor{}} / \Ex{\Tim{}}$ may be unbounded:
\newcommand{\stmtpropseneta}{
 Consider the family of branching processes with transitions $X \btran{p} \< X X \>$ and $X \btran{1-p} \bot$, where $0 < p < 1/2$.
 Then the ratio $\Ex{\Wor{X}} / \Ex{\Tim{X}}$ is unbounded for $p \to 1/2$.
}
\begin{proposition} \label{prop:seneta}
 \stmtpropseneta
\end{proposition}

%% file: practical.tex
\section{Case Studies} \label{sec:practical}

We have implemented a prototype tool in the form of a Maple worksheet,
 which allows to compute some of the quantities from the previous section:
  the termination probabilities, and distributions and expectations of work and time.
In this section, we use our tool for two case studies%
\footnote{Available at {\scriptsize \texttt{http://www.comlab.ox.ac.uk/people/Stefan.Kiefer/case-studies.mws}}.}%
, which also illustrate how probabilistic parallel programs can be modelled with pSJSs.
We only deal with normalised pSJSs in this section.

\subsection{Divide and Conquer}

The pSJS model lends itself to analyse parallel divide-and-conquer programs.
For simplicity, we assume that the problem is already given as a binary tree,
 and solving it means traversing the tree and combining the results of the children.
Figure~\ref{fig:divCon} shows generic parallel code for such a problem.

\newcommand{\einr}{\hspace{4mm}}%
\newcommand{\einrr}{\hspace{4mm}\einr}%
\newcommand{\einrrr}{\hspace{4mm}\einrr}%
\begin{figure}
%\fbox
\iftechrep{}{\vspace{-8mm}}
{\parbox{\textwidth}{ \flushleft
    \textbf{function} divCon(node) \\
    \einr \textbf{if} node.leaf() \textbf{then return} node.val() \\
    \einr \textbf{else parallel}  $\<$ val1 := divCon(node.c1), val2 := divCon(node.c2) $\>$ \\
    \hspace{9.2mm} \textbf{return} combine(val1, val2)
}}
\caption{A generic parallel divide-and-conquer program.}
\label{fig:divCon}
\end{figure}

\iftechrep{}{\vspace{-3mm}}
For an example, think of a routine for numerically approximating an integral $\int_{0}^1 f(x)\, dx$.
\newcommand{\osc}{\mathit{osc}}%
Given the integrand~$f$ and a subinterval~$I \subseteq [0,1]$, we assume that there is a function which computes $\osc_f(I) \in \N$,
 the ``oscillation'' of~$f$ in the interval~$I$,
 a measure for the need for further refinement.
If $\osc_f(I) = 0$, then the integration routine returns the approximation $1 \cdot f(1/2)$,
 otherwise it returns $I_1 + I_2$, where $I_1$ and $I_2$ are recursive approximations of $\int_{0}^{1/2} f(x)\, dx$ and $\int_{1/2}^1 f(x)\, dx$, respectively.%
\footnote{Such an adaptive approximation scheme is called ``local'' in~\cite{MalcolmSimpson75}.}

We analyse such a routine using probabilistic assumptions on the integrand:
Let $n, n_1, n_2$ be nonnegative integers such that $0 \le n_1{+}n_2 \le n$.
If $\osc_f([a,b]) = n$, then $\osc_f([a,(a{+}b)/2]) = n_1$ and $\osc_f([(a{+}b)/2,b]) = n_2$ with probability
 $x(n,n_1,n_2) := {n \choose n_1} \cdot {n - n_1 \choose n_2} \cdot \left(\frac{p}{2}\right)^{n_1} \cdot \left(\frac{p}{2}\right)^{n_2} \cdot (1{-}p)^{n-n_1-n_2}$,
  where $0 < p < 1$ is some parameter.\footnote{%
That means, the oscillation~$n$ in the interval $[a,b]$ can be thought of as distributed between $[a,(a+b)/2]$ and $[(a+b)/2,b]$ according
 to a ball-and-urn experiment, where each of the $n$ balls is placed in the $[a,(a{+}b)/2]$-urn and the $[(a{+}b)/2,b]$-urn with probability $p/2$, respectively,
  and in a trash urn with probability $1{-}p$.}
Of course, other distributions could be used as well.
The integration routine can then be modelled by the pSJS with $Q = \{q\}$ and $\Gamma = \{\<q q\>, 0, \ldots, n_{\max{}}\}$
 and the following rules:
\iftechrep{}{\vspace{-3mm}}
\begin{align*}
 0 \btran{1} q \quad \text{and} \quad \< q \ q \> \btran{1} q \quad \text{and} \quad n \btran{x(n,n_1,n_2)} \< n_1 \ n_2 \> \text{ for all $1 \le n \le n_{\max{}}$}\,,
\end{align*}
\iftechrep{}{\\[-6mm]}
 where $0 \le n_1{+}n_2 \le n$.
(Since we are merely interested in the performance of the algorithm, we can identify all return values with a single synchronisation state~$q$.)

Using our prototype, we computed $\Ex{\Wor{n}}$ and $\Ex{\Tim{n}}$ for $p=0.8$ and $n=0, 1, \ldots, 10$.
Figure~\ref{fig:graph-divCon} shows that $\Ex{\Wor{n}}$ increases faster with~$n$ than~$\Ex{\Tim{n}}$;
 i.e., the parallelism increases.
\begin{figure}
 \psfrag{0}{\hspace{1.5mm}$0$}
 \psfrag{1}{$1$}
 \psfrag{2}{$2$}
 \psfrag{3}{$3$}
 \psfrag{4}{$4$}
 \psfrag{5}{$5$}
 \psfrag{6}{$6$}
 \psfrag{7}{$7$}
 \psfrag{8}{$8$}
 \psfrag{9}{$9$}
 \psfrag{10}{$10$}
 \psfrag{20}{$20$}
 \psfrag{40}{$40$}
 \psfrag{60}{$60$}
 \psfrag{80}{$80$}
 \psfrag{tim}{$\Ex{\Tim{n}}$}
 \psfrag{wor}{$\Ex{\Wor{n}}$}
 \psfrag{n}{$n$}
 \includegraphics{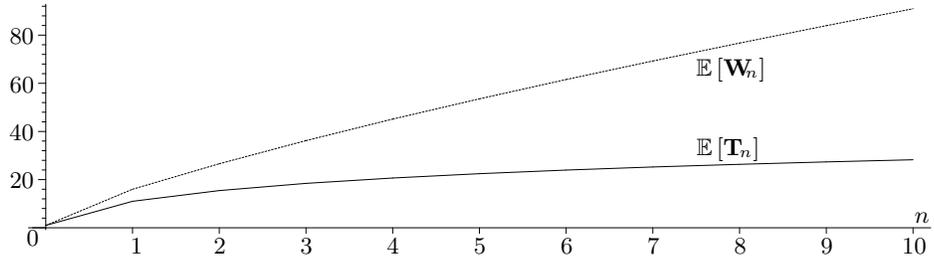}
\iftechrep{}{\vspace{-5mm}}
\caption{Expectations of time and work.}
\label{fig:graph-divCon}
\end{figure}

\subsection{Evaluation of Game Trees}

The evaluation of game trees is a central task of programs that are equipped with ``artificial intelligence'' to play games such as chess.
These game trees are min-max trees (see Figure~\ref{fig:game-tree}):
\newcommand{\xsc}{1.4}%
\begin{figure}
\iftechrep{}{\vspace{-5mm}}
\begin{tikzpicture}
[scale=1.0,nod/.style={circle,draw,inner sep=0pt,outer sep=0pt,minimum size=4mm}]
  \node (max)  at (\xsc*5,2) [nod,label={[label distance=-1mm]95:$3$}] {$\lor$};
  \node (min1) at (\xsc*2,1) [nod,label={[label distance=-1mm]95:$3$}] {$\land$};
  \node (min2) at (\xsc*5,1) [nod,label={[label distance=-1mm]95:${\le}3$}] {$\land$};
  \node (min3) at (\xsc*8,1) [nod,label={[label distance=-1mm]95:${\le}2$}] {$\land$};
  \draw (max)--(min1);
  \draw (max)--(min2);
  \draw (max)--(min3);
  \node (max1) at (\xsc*1,0) [nod,label={[label distance=-1mm]95:$3$}] {$\lor$};
  \node (max2) at (\xsc*2,0) [nod,label={[label distance=-1.5mm]95:${\ge}3$}] {$\lor$};
  \node (max3) at (\xsc*3,0) [nod,label={[label distance=-0.5mm]90:${\ge}4$}] {$\lor$};
  \node (max4) at (\xsc*4,0) [nod,label={[label distance=-1mm]95:$3$}] {$\lor$};
  \node (max5) at (\xsc*5,0) [nod] {$\lor$};
  \node (max6) at (\xsc*6,0) [nod] {$\lor$};
  \node (max7) at (\xsc*7,0) [nod,label={[label distance=-1mm]95:$2$}] {$\lor$};
  \node (max8) at (\xsc*8,0) [nod] {$\lor$};
  \node (max9) at (\xsc*9,0) [nod] {$\lor$};
  \draw (min1)--(max1);
  \draw (min1)--(max2);
  \draw (min1)--(max3);
  \draw (min2)--(max4);
  \draw (min2)--(max5);
  \draw (min2)--(max6);
  \draw (min3)--(max7);
  \draw (min3)--(max8);
  \draw (min3)--(max9);
  \node (leaf11) at (\xsc*2/3,-1)  {$3$};
  \node (leaf12) at (\xsc*3/3,-1)  {$2$};
  \node (leaf13) at (\xsc*4/3,-1)  {$2$};
  \node (leaf21) at (\xsc*5/3,-1)  {$3$};
  \node (leaf22) at (\xsc*6/3,-1)  {};
  \node (leaf23) at (\xsc*7/3,-1)  {};
  \node (leaf31) at (\xsc*8/3,-1)  {$4$};
  \node (leaf32) at (\xsc*9/3,-1)  {};
  \node (leaf33) at (\xsc*10/3,-1) {};
  \node (leaf41) at (\xsc*11/3,-1) {$3$};
  \node (leaf42) at (\xsc*12/3,-1) {$3$};
  \node (leaf43) at (\xsc*13/3,-1) {$1$};
  \node (leaf51) at (\xsc*14/3,-1) {};
  \node (leaf52) at (\xsc*15/3,-1) {};
  \node (leaf53) at (\xsc*16/3,-1) {};
  \node (leaf61) at (\xsc*17/3,-1) {};
  \node (leaf62) at (\xsc*18/3,-1) {};
  \node (leaf63) at (\xsc*19/3,-1) {};
  \node (leaf71) at (\xsc*20/3,-1) {$2$};
  \node (leaf72) at (\xsc*21/3,-1) {$1$};
  \node (leaf73) at (\xsc*22/3,-1) {$1$};
  \node (leaf81) at (\xsc*23/3,-1) {};
  \node (leaf82) at (\xsc*24/3,-1) {};
  \node (leaf83) at (\xsc*25/3,-1) {};
  \node (leaf91) at (\xsc*26/3,-1) {};
  \node (leaf92) at (\xsc*27/3,-1) {};
  \node (leaf93) at (\xsc*28/3,-1) {};
  \draw (max1)--(leaf11);
  \draw (max1)--(leaf12);
  \draw (max1)--(leaf13);
  \draw (max2)--(leaf21);
  \draw (max2)--(leaf22);
  \draw (max2)--(leaf23);
  \draw (max3)--(leaf31);
  \draw (max3)--(leaf32);
  \draw (max3)--(leaf33);
  \draw (max4)--(leaf41);
  \draw (max4)--(leaf42);
  \draw (max4)--(leaf43);
  \draw (max5)--(leaf51);
  \draw (max5)--(leaf52);
  \draw (max5)--(leaf53);
  \draw (max6)--(leaf61);
  \draw (max6)--(leaf62);
  \draw (max6)--(leaf63);
  \draw (max7)--(leaf71);
  \draw (max7)--(leaf72);
  \draw (max7)--(leaf73);
  \draw (max8)--(leaf81);
  \draw (max8)--(leaf82);
  \draw (max8)--(leaf83);
  \draw (max9)--(leaf91);
  \draw (max9)--(leaf92);
  \draw (max9)--(leaf93);
\end{tikzpicture}
\iftechrep{}{\vspace{-6mm}}
\caption{A game tree with value~$3$.} \label{fig:game-tree}
\end{figure}
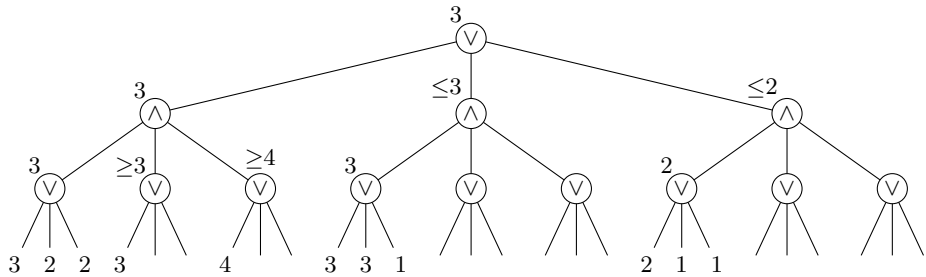
each node corresponds to a position of the game,
 and each edge from a parent to a child corresponds to a move that transforms the position represented by the parent to a child position.
Since the players have opposing objectives, the nodes alternate between max-nodes and min-nodes (denoted ${\lor}$ and ${\land}$, respectively).
A leaf of a game tree corresponds either to a final game position or to a position which is evaluated heuristically by the game-playing program;
 in both cases, the leaf is assigned a number.
Given such a leaf labelling, a number can be assigned to each node in the tree in the straightforward way;
 in particular, \emph{evaluating a tree} means computing the root value.

In the following, we assume for simplicity that each node is either a leaf or has exactly three children.
Figure~\ref{fig:parMax} shows a straightforward recursive parallel procedure for evaluating a max-node of a game tree.
(Of course, there is a symmetrical procedure for min-nodes.)
\begin{figure}
%\fbox
{\parbox{\textwidth}{ \flushleft
    \textbf{function} parMax(node) \\
    \einr \textbf{if} node.leaf() \textbf{then return} node.val() \\
    \einr \textbf{else parallel}  $\<$ val1 := parMin(node.c1), val2 := parMin(node.c2), val3 := parMin(node.c3) $\>$ \\
%    \hspace{26mm}                      val3 := parMin(node.c3) $\>$ \\
    \hspace{9.2mm} \textbf{return} $\max\{\text{val1, val2, val3}\}$
}}
\caption{A simple parallel program for evaluating a game tree.}
\label{fig:parMax}
\end{figure}

Notice that in Figure~\ref{fig:game-tree} the value of the root is~$3$, independent of some missing leaf values.
Game-playing programs aim at evaluating a tree as fast as possible, possibly by not evaluating nodes which are irrelevant for the root value.
The classic technique is called \emph{alpha-beta pruning}: it maintains an interval $[\alpha, \beta]$ in which the value of the current node is
 to be determined exactly.
If the value turns out to be below~$\alpha$ or above~$\beta$, it is safe to return $\alpha$ or~$\beta$, respectively.
This may be the case even before all children have been evaluated (a so-called \emph{cut-off}).
Figure~\ref{fig:seqMax} shows a sequential program for alpha-beta pruning, initially to be called ``seqMax(root, $-\infty$, $+\infty$)''.
Applying seqMax to the tree from Figure~\ref{fig:game-tree} results in several cut-offs: all non-labelled leaves are pruned.
\begin{figure}
%\fbox
\iftechrep{}{\vspace{-7mm}}
{\parbox{\textwidth}{ \flushleft
    \textbf{function} seqMax(node, $\alpha$, $\beta$) \\
    \einr \textbf{if} node.leaf() \textbf{then if} node.val() $\le \alpha$ \textbf{then return} $\alpha$ \\
    \hspace{27.5mm} \textbf{elsif} node.val() $\ge \beta$ \textbf{then return} $\beta$ \\
    \hspace{27.5mm} \textbf{else return} node.val() \\
    \einr \textbf{else} val1 := seqMin(node.c1, $\alpha$, $\beta$) \\
    \hspace{9.5mm}       \textbf{if} val1 $= \beta$ \textbf{then return} $\beta$ \\
    \hspace{9.5mm}       \textbf{else} val2 := seqMin(node.c2, val1, $\beta$) \\
    \hspace{15.8mm}                     \textbf{if} val2 $= \beta$ \textbf{then return} $\beta$ \\
    \hspace{15.8mm}                     \textbf{else return} seqMin(node.c3, val2, $\beta$)
}}
\caption{A sequential program for evaluating a game tree using alpha-beta pruning.}
\label{fig:seqMax}
\end{figure}

Although alpha-beta pruning may seem inherently sequential, parallel versions have been developed,
 often involving the \emph{Young Brothers Wait (YBW)} strategy~\cite{YBW89}.
It relies on a good ordering heuristic,
 i.e., a method that sorts the children of a max-node (resp.\ min-node) in increasing (resp.\ decreasing) order,
  without actually evaluating the children.
Such an ordering heuristic is often available, but usually not perfect.
The tree in Figure~\ref{fig:game-tree} is ordered in this way.
If alpha-beta pruning is performed on such an ordered tree, then either all children of a node are evaluated or only the first one.
The YBW method first evaluates the first child only and hopes that this creates a cut-off or, at least, decreases the interval $[\alpha, \beta]$.
If the first child fails to cause a cut-off, YBW \emph{speculates} that both ``younger brothers`` need to be evaluated,
 which can be done in parallel without wasting work.
A wrong speculation may affect the performance, but not the correctness.
Figure~\ref{fig:YBWMax} shows a YBW-based program.
Similar code is given in~\cite{YBWCilk02} using \emph{Cilk}, a C-based parallel programming language.
\begin{figure}
%\fbox
{\parbox{\textwidth}{ \flushleft
    \textbf{function} YBWMax(node, $\alpha$, $\beta$) \\
    \einr \textbf{if} node.leaf() \textbf{then if} node.val() $\le \alpha$ \textbf{then return} $\alpha$ \\
    \hspace{27.5mm} \textbf{elsif} node.val() $\ge \beta$ \textbf{then return} $\beta$ \\
    \hspace{27.5mm} \textbf{else return} node.val() \\
    \einr \textbf{else} val1 := YBWMin(node.c1, $\alpha$, $\beta$) \\
    \hspace{9.5mm}       \textbf{if} val1 $= \beta$ \textbf{then return} $\beta$ \\
    \hspace{9.5mm}       \textbf{else parallel} $\<$ val2 := YBWMin(node.c2, val1, $\beta$), val3 := YBWMin(node.c3, val1, $\beta$) $\>$ \\
%    \hspace{33mm}                                   val3 := YBWMin(node.c3, val1, $\beta$) $\>$ \\
    \hspace{15mm}                     \textbf{return} $\max\{\text{val2, val3}\}$ \\
}}
\caption{A parallel program based on YBW for evaluating a game tree.}
\label{fig:YBWMax}
\end{figure}

We evaluate the performance of these three (deterministic) programs using probabilistic assumptions
 about the game trees.
More precisely, we assume the following:
 Each node has exactly three children with probability~$p$, and is a leaf with probability~$1{-}p$.
 A leaf (and hence any node) takes as value a number from $\N_4 := \{0,1,2,3,4\}$, according to a distribution described below.
 In order to model an ordering heuristic on the children, each node carries a parameter~$e \in \N_4$ which intuitively corresponds to its expected value.
 If a max-node with parameter~$e$ has children, then they are min-nodes with parameters $e$, $e{\ominus}1$, $e{\ominus}2$, respectively,
  where $a{\ominus}b := \max\{a{-}b, 0\}$;
 similarly, the children of a min-node with parameter~$e$ are max-nodes with parameters $e$, $e{\oplus}1$, $e{\oplus}2$,
  where $a{\oplus}b := \min\{a{+}b, 4\}$.
 A leaf-node with parameter~$e$ takes value~$k$ with probability ${4 \choose k} \cdot (e/4)^k \cdot (1{-}e/4)^{4-k}$;
  i.e., a leaf value is binomially distributed with expectation~$e$.
\newcommand{\Max}{\mathit{Max}}%
\newcommand{\Min}{\mathit{Min}}%
 One could think of a game tree as the terminal tree of a branching process with
  $\Gamma = \{\Max(e), \Min(e) \mid e \in \{0, \ldots, 4\}\}$ and $Q = \N_4$ and the rules
  $\Max(e) \btran{p} \< \Min(e) \ \Min(e{\ominus}1) \ \Min(e{\ominus}2) \>$ and  $\Max(e) \btran{x(k)} k$,
 with $x(k) := (1{-}p) \cdot {4 \choose k} \cdot (e/4)^k \cdot (1{-}e/4)^{4-k}$ for all $e, k \in \N_4$,
  and similar rules for~$\Min(e)$.

\newcommand{\MaTwo}{\Max 2}%
We model the YBW-program from Figure~\ref{fig:YBWMax} running on such random game trees by the pSJS with
 $Q = \{0,1,2,3,4,q({\lor}),q({\land})\} \cup \{ q(\alpha,\beta,{\lor},e), q(\alpha,\beta,{\land},e) \mid 0 \le \alpha < \beta \le 4, \ 0 \le e \le 4 \}
  \cup \{q(a,b) \mid 0 \le a, b \le 4\}$
 and the following rules:
  \begin{align*} &
   \Max(\alpha,\beta,e) \btran{x(0) + \cdots + x(\alpha)} \alpha, \quad
   \Max(\alpha,\beta,e) \btran{x(\beta) + \cdots + x(4)}  \beta , \quad
   \Max(\alpha,\beta,e) \btran{x(k)} k \\&
   \Max(\alpha,\beta,e) \btran{p} \< \Min(\alpha,\beta,e) \ \ q(\alpha,\beta,{\lor},e{\ominus}1) \> \\&
   \< \beta \ \ q(\alpha,\beta,{\lor},e) \> \btran{1} \beta, \qquad
   \< \gamma \ \ q(\alpha,\beta,{\lor},e) \> \btran{1} \< \MaTwo(\gamma,\beta,e) \ \ q({\lor}) \> \\&
   \MaTwo(\alpha,\beta,e) \btran{1} \< \Min(\alpha,\beta,e) \ \ \Min(\alpha,\beta,e{\ominus}1) \> \\&
   \< a \ \ b \> \btran{1} q(a,b), \qquad \< q(a,b) \ \ q({\lor}) \> \btran{1} \max\{a,b\} \,,
  \end{align*}
 where $0 \le \alpha \le \gamma < \beta \le 4$ and $\alpha < k < \beta$ and $0 \le e \le 4$ and $0 \le a, b \le 4$.
There are analogous rules with $\Min$ and~$\Max$ exchanged.
Notice that the rules closely follow the program from Figure~\ref{fig:YBWMax}.
The programs parMax and seqMax from Figures~\ref{fig:parMax} and~\ref{fig:seqMax} can be modelled similarly.

\newcommand{\YBW}{\text{YBW}}%
\newcommand{\pa}{\text{par}}%
\newcommand{\se}{\text{seq}}%
Let $T(\YBW,p) := \Ex{\Tim{\Max(0,4,2)} \mid \tRun{\Max(0,4,2)}{2}}$;
 i.e., $T(\YBW,p)$ is the expected time of the YBW-program called with a tree with value~$2$ and whose root is a max-node with parameter~$2$.
(Recall that $p$ is the probability that a node has children.)
Let $W(\YBW,p)$ defined similarly for the expected work, and define these numbers also for par and seq instead of~\YBW,
 i.e., for the programs from Figures~\ref{fig:parMax} and~\ref{fig:seqMax}.
Using our prototype we computed $W(\se,p) = 1.00, 1.43, 1.96, 2.63, 3.50, 4.68, 6.33$ for $p = 0.00, 0.05, 0.10, 0.15, 0.20, 0.25, 0.30$.
Since the program seq is sequential, we have the same sequence for $T(\se,p)$.
To assess the speed of the parallel programs par and YBW,
 we also computed the percentaged increase of their runtime relative to seq, i.e., $100 \cdot (T(\pa,p)/T(\se,p) - 1)$, and similarly for~YBW.
Figure~\ref{fig:graph-runtime} shows the results.
\begin{figure}
\iftechrep{}{\vspace{-6mm}}
 \psfrag{par}{\pa}
 \psfrag{YBW}{\YBW}
 \psfrag{p}{$p$}
 \psfrag{0.05}{$0.05$}
 \psfrag{0.1}{$0.1$}
 \psfrag{0.15}{$0.15$}
 \psfrag{0.2}{$0.2$}
 \psfrag{0.25}{$0.25$}
 \psfrag{0.3}{$0.3$}
 \psfrag{\26110}{$\hspace{-3mm}-10\%$}
 \psfrag{0}{}
 \psfrag{10}{$\hspace{-5mm}+10\%$}
 \psfrag{20}{$\hspace{-5mm}+20\%$}
 \psfrag{30}{$\hspace{-5mm}+30\%$}
 \psfrag{40}{$\hspace{-5mm}+40\%$}
 \psfrag{50}{$\hspace{-5mm}+50\%$}
 \includegraphics{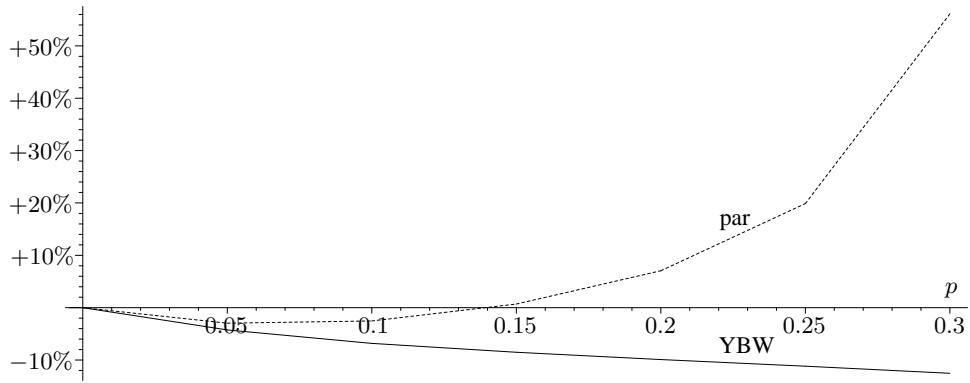}
\iftechrep{}{\vspace{-3mm}}
\caption{Percentaged runtime increase of par and YBW relative to seq.}
\label{fig:graph-runtime}
\end{figure}
One can observe that for small values of~$p$ (i.e., small trees), the program par is slightly faster than seq because of its parallelism.
For larger values of~$p$, par still evaluates all nodes in the tree, whereas seq increasingly benefits from cut-offs of potentially deep branches.
Using Proposition~\ref{prop:bp-critical},
 one can prove $W(\pa,\frac13) = T(\pa,\frac13) = \infty > W(\se,\frac13)$.%
\footnote{In fact, $W(\se,p)$ is finite even for values of~$p$ which are slightly larger than~$\frac13$; in other words, seq cuts off infinite branches.}
The figure also shows that the YBW-program is faster than seq: the advantage of YBW increases with~$p$ up to about $10\%$.

We also compared the \emph{work} of YBW with seq, and found that the percentaged increase
%.153533e-1 .542288e-1 .1122159 .1890123 .2864358 .4079437
ranges from~$0$ to about $+0.4\%$ for $p$ between $0$ and~$0.3$.
This means that YBW wastes almost no work;
 in other words, a sequential version of YBW would be almost as fast as seq.
An interpretation is that the second child rarely causes large cut-offs.
Of course, all of these findings could depend on the exact probabilistic assumptions on the game trees.

%% file: conclusions.tex
\section{Conclusions and Future Work} \label{sec:conclusions}

We have introduced pSJSs, a model for probabilistic parallel programs with process spawning and synchronisation.
We have studied the basic performance measures of termination probability, space, work, and time.
In our results the upper complexity bounds coincide with the best ones known for pPDSs,
 and the lower bounds also hold for pPDSs.
This suggests that analysing pSJSs is no more expensive than analysing pPDSs.
The pSJS model is amenable to a practical performance analysis.
Our two case studies have demonstrated the modelling power of pSJSs:
one can use pSJSs to model, analyse, and compare the performance of parallel programs under probabilistic assumptions.

We intend to develop model-checking algorithms for pSJSs.
It seems to us that a meaningful functional analysis should not only model-check the Markov chain induced by the pSJS,
 but rather take the individual process ``histories'' into account.

%% file: app-pPDS.tex
\section{Proofs of Section~\ref{sub:rel-pPDS}}

%\subsection{Proof of Proposition~\ref{prop:serialisation}}

Here is a restatement of Proposition~\ref{prop:serialisation}.

\begin{qproposition}{\ref{prop:serialisation}}
 \stmtpropserialisation
\end{qproposition}

\begin{proof}
 We define a bijection $b : \tRunM{M_S}{\sigma}{q} \to \tRunM{M_{S_P}}{\Box \sigma}{\overline{q}}$.
 Since these sets only contain runs that reach a terminal state (namely, $q$ and~$\overline{q}$, respectively),
  we identify in this proof a run with the (finite) path that leads to the terminal state.
 For a path~$w$ in~$M_{S_P}$ with length~$n$ we write $w \rfloor \sigma$ for the path $z$ of length~$n$ with $z(i) = w(i) \sigma$ for all $0 \le i \le n-1$.

 Let $w \in \tRunM{M_S}{\sigma}{q}$.
 We define~$b(w)$ inductively by order of the length~$n$ of~$w$.
 The run~$w$ has one of the following forms:
 \begin{itemize}
  \item
   Let $w = w(0)$, where $w(0) = \sigma = q$.
    Then we set $b(w) := \Box q,\ \overline{q}$.
  \item
   Let $w = a,\ \<\sigma_1 \sigma_2\>, \ \ldots,\ w(k),\ \ldots,\ q$, where $a \in \Gamma$ and $w(k)= \<q_1 q_2\> $ for some~$k$ with $1 \le k \le n-2$,
    such that $w(i) \not\in \Gamma$ for $i \in \{1, \ldots, k-1\}$.
   Consider the left and the right subtree of the root nodes in the run between $w(1) = \<\sigma_1 \sigma_2\>$ and~$w(k) = \<q_1 q_2\>$.
   By the semantics of pSJSs, there are corresponding runs $u_1 = \sigma_1,\ \ldots,\ q_1$ and $u_2 = \sigma_2,\ \ldots,\ q_2$
    such that $u_1 \in \tRunM{M_S}{\sigma_1}{q_1}$ and $u_2 \in \tRunM{M_S}{\sigma_2}{q_2}$ and $\max\{|u_1|, |u_2|\} = k$.
   Let $z_1 := b(u_1)$ and $z_2 := b(u_2)$ and $z_3 := b(w(k),\ \ldots,\ w(n-1))$.
   Then we set $b(w) := \Box a,\ z_1 \rfloor \sigma_2,\ z_2 \rfloor \widetilde{q_1},\ z_3$.
  \item
   Let $w = a,\ \sigma_1,\ \ldots,\ q$, where $a \in \Gamma$ and $\sigma_1 \in \Sigma \setminus \<Q Q\>$.
   Then we set $b(w) := \Box a,\ b(\sigma_1,\ \ldots,\ q)$.
 \end{itemize}
 Notice that in all cases $b(w) \in \tRunM{M_{S_P}}{\Box \sigma}{\overline{q}}$.
 It is easy to check that $b$ is a bijection.
 It is also easy to see that $\P{\{w\}} = \P{\{b(w)\}}$, because in both cases the probability is the product of the probabilities of the applied
  transition rules (of course, taking multiplicities into account).
\qed
\end{proof}

%% file: app-termination.tex
\section{Proofs of Section~\ref{sub:termination-probabilities}}

\subsection{Proof of Proposition~\ref{prop:termination}}

Here is a restatement of Proposition~\ref{prop:termination}.

\begin{qproposition}{\ref{prop:termination}}
 \stmtproptermination
\end{qproposition}

\begin{proof}
 The proposition can be proved by adapting the corresponding proofs for pPDSs from \cite{EKM04} or~\cite{EYstacs05Extended}.
 Alternatively, we can use the ``serialisation'' procedure from Section~\ref{sub:rel-pPDS}
  and the equality $\term{\sigma}{q} = \term{\Box \sigma}{\overline{q}}$,
  reducing the problem from pSJSs to pPDSs.
 Then we can take the equation system from \cite{EKM04,EYstacs05Extended} for $\term{\Box \sigma}{\overline{q}}$ and
  compress it by substituting variables with the right-hand side of their equation.
 This gives the same equation system as the one above.
\qed
\end{proof}

\subsection{Proof of Theorem~\ref{thm:termination-probabilities}}

Here is a restatement of Theorem~\ref{thm:termination-probabilities}.

\begin{qtheorem}{\ref{thm:termination-probabilities}}
 \stmtthmterminationprobabilities
\end{qtheorem}
\begin{proof}
 The respective claims are shown for pPDSs in~\cite{EtessamiY05,EYstacs05Extended}.
 For statements (1) and~(2), we use Proposition~\ref{prop:serialisation} to reduce the problem to pPDSs.
 For statement~(3), recall from Section~\ref{sub:rel-pPDS} that pPDSs can be encoded as pSJSs.
\qed
\end{proof}

%% file: app-finite-space.tex
\section{Proofs of Section~\ref{sub:finite-space}}

\subsection{Proof of Lemma~\ref{lem:unbounded-poly-time}}

Here is a restatement of Lemma~\ref{lem:unbounded-poly-time}.
\begin{qlemma}{\ref{lem:unbounded-poly-time}}
 \stmtlemunboundedpolytime
\end{qlemma}
\begin{proof}
 We write $t_1 \tran{}^* t_2$ if $t_2$ can be reached from~$t_1$ in the Markov chain~$M_S$ induced by the pSJS~$S$;
  i.e., $\mathord{\tran{}^*}$ is the reflexive and transitive closure of~$\mathord{\tran{}}$.
 Define $\mathord{\Rightarrow} := \mathord{\tran{}^*} \cap ( \Sigma \times \Sigma )$;
  i.e., we have $\sigma_1 \Rightarrow \sigma_2$ if and only if $\sigma_2 \in \Sigma$ can be reached from $\sigma_1 \in \Sigma$.
 The relation~$\mathord{\Rightarrow}$ can be computed in polynomial time
  using the fact that it is the smallest subset of $\Sigma \times \Sigma$ that satisfies:
  \begin{itemize}
   \item $\sigma \Rightarrow \sigma$ for all $\sigma \in \Sigma$;
   \item $\sigma_1 \btran{} \sigma_2 \Rightarrow \sigma_3$ implies $\sigma_1 \Rightarrow \sigma_3$;
   \item $\sigma_1 \btran{} \<\sigma_2 \sigma_3\>$ and $\term{\sigma_2}{q_2} > 0$ and $\term{\sigma_3}{q_3} > 0$
        and $\<q_2 q_3\> \Rightarrow \sigma_4$  imply $\sigma_1 \Rightarrow \sigma_4$.
  \end{itemize}
 For a tree $t \in T(\Sigma)$, let $h(t)$ denote its \emph{height}, i.e., the maximal distance of a leaf to the root.
 Moreover, we define for each $k \in \N$:
 \[
  U_k := \{ \sigma \in \Sigma \mid \text{there is a tree $t \in T(\sigma)$ with $\sigma \tran{}^* t$ and $h(t) \ge k$} \} \,.
 \]
 Notice that $\Sigma = U_0 \supseteq U_1 \supseteq U_2 \ldots$
 It is easy to see that we have for all $k \in \N$:
 \[
  U_{k+1} = \{a \in \Gamma \mid \exists \text{ transition } b \btran{} \<\sigma_1 \sigma_2\> :
     a \Rightarrow b \text{ and } \{\sigma_1, \sigma_2\} \cap U_k \ne \emptyset \} \,.
 \]
 It follows from this characterisation that the sequence $U_0 \supseteq U_1 \supseteq U_2 \ldots$ stabilises after at most $|\Sigma|$ steps;
  i.e., we have $U_{|\Sigma|} = U_{|\Sigma|+1} = \ldots$
 In other words, we have:
 \[
  U_{|\Sigma|} = \{ a \in \Gamma \mid \forall n \in \N : \exists t \in T(\Sigma) : a \tran{}^* t \text{ and } h(t) > n \} \,.
 \]
 For each tree $t \in T(\Sigma)$ we have $h(t) \le |t| \le 2^{h(t)}$
  (recall that $|t|$ is the length of~$t$ not counting the symbols `$\<$' and `$\>$').
 Therefore, we have $U = U_{|\Sigma|}$, so it suffices to compute~$U_{|\Sigma|}$, which can be done in polynomial time
  with the above characterisation.
\qed
\end{proof}

\subsection{Proof of Proposition~\ref{prop:finite-space}}

Here is a restatement of Proposition~\ref{prop:finite-space}.

\begin{qproposition}{\ref{prop:finite-space}}
 \stmtpropfinitespace
\end{qproposition}
\begin{proof}
 Statement~(1) follows from the fact that $\barS$ has, by construction,
  for all $q_1, q_2 \in \barQ$ a transition with $\<q_1,q_2\>$ on the left hand side.
 (In other words, $\barGamma \supseteq \barQ \times \barQ$.)

 For statement~(2) observe that $\barS$ is obtained from~$S$ via two steps: a normalisation step,
  and a step where transitions with process symbols $b \in B$ on the left hand side are replaced.
 We argue that neither of those steps modifies the value $\term{a}{q}$ for $a \in \Gamma$ and $q \in Q$.
 The first step does not modify~$\term{a}{q}$, because the normalisation affects only those runs
  that would not terminate in a synchronisation state without normalisation.
 With normalisation, those runs may terminate in the new synchronisation state introduced by the normalisation, but not in~$q$.
 The second step does not modify~$\term{a}{q}$, because it affects only those runs
  that would otherwise not terminate.
 With the modification in the second step, those runs may terminate in~$\overline{q}$, but not in~$q$.

\newcommand{\Sn}{S^\text{norm}}
\newcommand{\Ss}{S^\text{stat}}
\newcommand{\Good}{\mathit{Good}}
 For statement~(3), it is convenient to consider a version of~$S$ ``in between'' the first and the second modification step.
 More precisely, let $\Sn$ denote the pSJS after the normalisation step,
  and let $\Ss$ denote the pSJS obtained from~$\Sn$ by removing all transitions with symbols $b \in B$ on the left hand side
  and replacing them with a transition $b \btran{1} b$.
 Notice that $\P{\Spa{a} < \infty = \Tim{a}}$ is the same in~$S$ and~$\Sn$ and~$\Ss$ for all $a \in \Gamma$.
 Denote by $\Good(a)$ the set of those runs $w \in \Run[M_{\Ss}](a)$ that reach a bottom strongly connected component (BSCC) of~$M_{\Ss}$.
 For any $n \in \N$ there are only finitely many trees~$t$ of~$M_{\Ss}$ such that $|t| \le n$.
 Hence it follows using standard arguments on finite Markov chains that in~$\Ss$ we have for all $n \in \N$
 \begin{align}
  \P{\Spa{a} \le n} & = \P{\Spa{a} \le n \text{ and } \Good(a)} && \text{, and so} \notag \\
  \P{\Spa{a} \le n \text{ and } \Tim{a}=\infty} & = \P{\Spa{a} \le n \text{ and } \Tim{a}=\infty \text{ and } \Good(a)} && \text{, and so} \notag \\
  \P{\Spa{a} < \infty = \Tim{a}} & = \P{\Spa{a} < \infty = \Tim{a} \text{ and } \Good(a)} \,. \label{eq:prop-finite-space}
 \end{align}
 Observe that each BSCC of~$M_{\Ss}$ consists of exactly one tree~$t$ such that $\Front(t)$ is either empty or consists only of elements of~$B$.
 So there is a natural 1-to-1 correspondence between the runs of~$\Ss$ that satisfy ``$\Spa{a} < \infty = \Tim{a} \text{ and } \Good(a)$''
  and the runs of~$\barS$ that are in~$\tRun{a}{\overline{q}}$.
 With~\eqref{eq:prop-finite-space} we get $\P{\Spa{a} < \infty = \Tim{a}} = \term{a}{\overline{q}}$.
\qed
\end{proof}

\subsection{Proof of Theorem~\ref{thm:finite-space}}

Here is a restatement of Theorem~\ref{thm:finite-space}.

\begin{qtheorem}{\ref{thm:finite-space}}
 \stmtthmfinitespace
\end{qtheorem}
\begin{proof}
 It follows from Proposition~\ref{prop:finite-space} that $s = \sum_{q \in \barQ} \term{a}{q}$,
  hence $s$ is expressible by Theorem~\ref{thm:termination-probabilities}~(1).
 Moreover, it follows using Theorem~\ref{thm:termination-probabilities}~(2) that deciding whether $s=0$ is in~\PTime.
 It remains to show statement~(3).

 We draw from a reduction in~\cite{EYstacs05Extended}, where it is shown that deciding if a pPDS terminates with probability~$1$ is \PosSLP-hard.
 As a gadget for that,
  Etessami and Yannakakis~\cite{EYstacs05Extended} compute, given a \PosSLP\ instance, a pPDS $S$ with the following properties.
 (More precisely, they construct an equivalent recursive Markov chain.)
 The starting configuration is $qa$, and after having left the initial configuration,
  $S$ reaches a configuration of the form $q \alpha$ with $\alpha \in \Gamma^*$ again with probability~$1$.
 At that time, the configuration is~$qaa$ with some probability~$p$, and $q$ with probability $1-p$.
 Moreover, the time (and hence space) needed to reach either of those configurations
  is essentially bounded by the size of the given \PosSLP\ instance, so it is finite.
 Furthermore, the given \PosSLP\ instance is a ``yes instance'' if and only if $p > \frac12$.
 It is easy to see that $\P{\Spa{qa} = \infty} > 0$ in~$S$
  if and only if $\P{\Spa{qa} = \infty} > 0$ in the pPDS~$\barS$ that consists only of the transitions
  $qa \btran{p} qaa$ and $qa \btran{1-p} q$.
 The Markov chain induced by~$\barS$, in turn, is isomorphic to the simple random walk $X_0, X_1, \ldots$ on~$\N$ with
  $X_0 = 1$ and $\P{X_{i+1} = 0 \mid X_{i} = 0} = 1$ and
 \begin{align*}
   \P{X_{i+1} = n+1 \mid X_{i} = n > 0} & = p \qquad \text{ and } \\
   \P{X_{i+1} = n-1 \mid X_{i} = n > 0} &= 1-p \,.
 \end{align*}
 It is well-known (see e.g.~\cite{Spitzer:book}) that for this random walk we have $\P{\sup_{i \in \N} X_i = \infty} > 0$ if and only if $p > \frac12$.
 It follows that we have $\P{\Spa{qa} < \infty} < 1$ in~$S$ if and only if $p > \frac12$.
\qed
\end{proof}

%% file: app-work-time.tex
\section{Proofs of Section~\ref{sub:work-time}}

\subsection{Proof of Proposition~\ref{prop:transformation}}

Here is a restatement of Proposition~\ref{prop:transformation}.

\begin{qproposition}{\ref{prop:transformation}}
 \stmtproptransformation
\end{qproposition}

\begin{proof}
 Define, for $q \in Q$ and $\sigma \in \Sigma$,
  \begin{align*}
   D_{\sigma q}(n)      & := \P{\tRun{\sigma}{q}, \ \Wor{\sigma} = n \mid \Run(\sigma)} \quad \text{and} \\
   D_{\bs{\sigma q}}(n) & := \P{\Wor{\bs{\sigma q}} = n \mid \Run(\bs{\sigma q})} \,.
  \end{align*}
 Notice that $\bs{\sigma q}$ and thus $D_{\bs{\sigma q}}(n)$ are undefined, if and only if $\term{\sigma}{q} = 0$.
 For the arithmetical expressions in the rest of this proof, we assume $0 \cdot \mathit{undefined} = 0$.
 For the statement on~$\Wor{}$ in the proposition,
  it suffices to show $D_{\sigma q}(n) = \term{\sigma}{q} \cdot D_{\bs{\sigma q}}(n)$ for $n \ge 0$.
 We proceed by induction on~$n$.

 Let $n=0$.
 If $\term{\sigma}{q} = 0$, then $D_{\sigma q}(0) = 0 = 0 \cdot \mathit{undefined} = \term{\sigma}{q} \cdot D_{\bs{\sigma q}}(0)$.
 If $\term{\sigma}{q} > 0$ and $\sigma = q$,
  then $D_{\sigma q}(0) = 1 = 1 \cdot 1 = \term{\sigma}{q} \cdot D_{\bot}(0) = \term{\sigma}{q} \cdot D_{\bs{\sigma q}}(0)$.
 If $\term{\sigma}{q} > 0$ and $\sigma \in \Gamma$, then $D_{\sigma q}(0) = 0 = \term{\sigma}{q} \cdot 0 = \term{\sigma}{q} \cdot D_{\bs{\sigma q}}(0)$.

 Let $n > 0$.
 If $\term{\sigma}{q} = 0$, then $D_{\sigma q}(n) = 0 = 0 \cdot \mathit{undefined} = \term{\sigma}{q} \cdot D_{\bs{\sigma q}}(n)$.
 If $\term{\sigma}{q} > 0$ and $\sigma = q$, then $D_{\sigma q}(n) = 0 = 1 \cdot 0 = \term{\sigma}{q} \cdot D_{\bs{\sigma q}}(n)$.
 If $\term{\sigma}{q} > 0$ and $\sigma = a \in \Gamma$, then
 \begin{align*}
  D_{a q}(n)
   & = \mathop{\sum_{a \btran{p} \< \sigma_1 \sigma_2 \>}}_{\<q_1 q_2\> \in \Gamma \cap \<Q Q\>}
       \mathop{\sum_{i,j,k \in \N}}_{1+i+j+k=n}
         p \cdot D_{\sigma_1 q_1}(i) \cdot D_{\sigma_2 q_2}(j) \cdot D_{\<q_1 q_2\> q}(k) \\
   & \qquad + \mathop{\sum_{a \btran{p} \sigma'}}_{\sigma' \in \Sigma \setminus \<Q Q\>} p \cdot D_{\sigma' q}(n-1) \\
   & = \mathop{\sum_{a \btran{p} \< \sigma_1 \sigma_2 \>}}_{\<q_1 q_2\> \in \Gamma \cap \<Q Q\>}
       \mathop{\sum_{i,j,k \in \N}}_{1+i+j+k=n}
         p \cdot \term{\sigma_1}{q_1} \cdot \term{\sigma_2}{q_2} \cdot \term{\<q_1 q_2\>}{q} \cdot \mbox{} \\
   & \hspace{50mm}       \mbox{} \cdot D_{\bs{\sigma_1 q_1}}(i) \cdot D_{\bs{\sigma_2 q_2}}(j) \cdot D_{\bs{\<q_1 q_2\> q}}(k) \\
   & \qquad + \mathop{\sum_{a \btran{p} \sigma'}}_{\sigma' \in \Sigma \setminus \<Q Q\>} p \cdot \term{\sigma'}{q} \cdot D_{\bs{\sigma' q}}(n-1) \\
   & = \term{a}{q} \cdot \left( \sum_{\bs{a q} \btran{p'} \< \bs{\sigma_1 q_1} \bs{\sigma_2 q_2} \bs{\<q_1 q_2\> q} \>}
                                \mathop{\sum_{i,j,k \in \N}}_{1+i+j+k=n}
                                   p' \cdot D_{\bs{\sigma_1 q_1}}(i) \cdot D_{\bs{\sigma_2 q_2}}(j) \cdot D_{\bs{\<q_1 q_2\> q}}(k) \right. \\
   & \qquad \qquad \qquad \left. + \sum_{\bs{a q} \btran{p'} \bs{\sigma' q}} p' \cdot D_{\bs{\sigma' q}}(n-1) \right) \\
   & = \term{a}{q} \cdot D_{\bs{a q}}(n) \,,
 \end{align*}
 where the first and the last equality are by the definition of~$\Wor{}$ and the semantics of pSJSs,
  the second equality is by the induction hypothesis, and the third equality is by the definition of~$\barGamma$.
 This proves the statement on~$\Wor{}$ in the proposition.

 The proof of the statement on~$\Tim{}$ is similar.
 Define, for $q \in Q$ and $\sigma \in \Sigma$,
  \begin{align*}
   E_{\sigma q}(n)      & := \P{\tRun{\sigma}{q}, \ \Tim{\sigma} \le n \mid \Run(\sigma)} \quad \text{and} \\
   E^=_{\sigma q}(n)    & := \P{\tRun{\sigma}{q}, \ \Tim{\sigma} = n \mid \Run(\sigma)} \quad \text{and} \\
   E_{\bs{\sigma q}}(n) & := \P{\Tim{\bs{\sigma q}} \le n \mid \Run(\bs{\sigma q})} \,.
  \end{align*}
 Notice that $\bs{\sigma q}$ and thus $E_{\bs{\sigma q}}(n)$ are undefined, if and only if $\term{\sigma}{q} = 0$.
 It suffices to show $E_{\sigma q}(n) \le \term{\sigma}{q} \cdot E_{\bs{\sigma q}}(n)$ for $n \ge 0$.
 We proceed by induction on~$n$.

 Let $n=0$.
 If $\term{\sigma}{q} = 0$, then $E_{\sigma q}(0) = 0 = 0 \cdot \mathit{undefined} = \term{\sigma}{q} \cdot E_{\bs{\sigma q}}(0)$.
 If $\term{\sigma}{q} > 0$ and $\sigma = q$,
  then $E_{\sigma q}(0) = 1 = 1 \cdot 1 = \term{\sigma}{q} \cdot E_{\bot}(0) = \term{\sigma}{q} \cdot E_{\bs{\sigma q}}(0)$.
 If $\term{\sigma}{q} > 0$ and $\sigma \in \Gamma$, then $E_{\sigma q}(0) = 0 = \term{\sigma}{q} \cdot 0 = \term{\sigma}{q} \cdot E_{\bs{\sigma q}}(0)$.

 Let $n > 0$.
 If $\term{\sigma}{q} = 0$, then $E_{\sigma q}(n) = 0 = 0 \cdot \mathit{undefined} = \term{\sigma}{q} \cdot E_{\bs{\sigma q}}(n)$.
 If $\term{\sigma}{q} > 0$ and $\sigma = q$, then $E_{\sigma q}(n) = 1 = 1 \cdot 1 = \term{\sigma}{q} \cdot E_{\bs{\sigma q}}(n)$.
 If $\term{\sigma}{q} > 0$ and $\sigma = a \in \Gamma$, then
 \begin{align*}
  E_{a q}(n)
   & = \mathop{\sum_{a \btran{p} \< \sigma_1 \sigma_2 \>}}_{\<q_1 q_2\> \in \Gamma \cap \<Q Q\>}
       \mathop{\sum_{i,j,k \in \N}}_{1+\max\{i,j\}+k \le n}
         p \cdot E^=_{\sigma_1 q_1}(i) \cdot E^=_{\sigma_2 q_2}(j) \cdot E^=_{\<q_1 q_2\> q}(k) \\
   & \qquad + \mathop{\sum_{a \btran{p} \sigma'}}_{\sigma' \in \Sigma \setminus \<Q Q\>} p \cdot E_{\sigma' q}(n-1) \\
   & \le \mathop{\sum_{a \btran{p} \< \sigma_1 \sigma_2 \>}}_{\<q_1 q_2\> \in \Gamma \cap \<Q Q\>}
       \mathop{\sum_{i,j,k \in \N}}_{1+\max\{i,j,k\} \le n}
         p \cdot E^=_{\sigma_1 q_1}(i) \cdot E^=_{\sigma_2 q_2}(j) \cdot E^=_{\<q_1 q_2\> q}(k) \\
   & \qquad + \mathop{\sum_{a \btran{p} \sigma'}}_{\sigma' \in \Sigma \setminus \<Q Q\>} p \cdot E_{\sigma' q}(n-1) \\
   & = \mathop{\sum_{a \btran{p} \< \sigma_1 \sigma_2 \>}}_{\<q_1 q_2\> \in \Gamma \cap \<Q Q\>}
         p \cdot E_{\sigma_1 q_1}(n-1) \cdot E_{\sigma_2 q_2}(n-1) \cdot E_{\<q_1 q_2\> q}(n-1) \\
   & \qquad + \mathop{\sum_{a \btran{p} \sigma'}}_{\sigma' \in \Sigma \setminus \<Q Q\>} p \cdot E_{\sigma' q}(n-1) \\
   & \le \mathop{\sum_{a \btran{p} \< \sigma_1 \sigma_2 \>}}_{\<q_1 q_2\> \in \Gamma \cap \<Q Q\>}
         p \cdot \term{\sigma_1}{q_1} \cdot \term{\sigma_2}{q_2} \cdot \term{\<q_1 q_2\>}{q} \cdot \mbox{} \\
   & \hspace{50mm}       \mbox{} \cdot E_{\bs{\sigma_1 q_1}}(n-1) \cdot E_{\bs{\sigma_2 q_2}}(n-1) \cdot E_{\bs{\<q_1 q_2\> q}}(n-1) \\
   & \qquad + \mathop{\sum_{a \btran{p} \sigma'}}_{\sigma' \in \Sigma \setminus \<Q Q\>} p \cdot \term{\sigma'}{q} \cdot E_{\bs{\sigma' q}}(n-1) \\
   & = \term{a}{q} \cdot \left( \sum_{\bs{a q} \btran{p'} \< \bs{\sigma_1 q_1} \bs{\sigma_2 q_2} \bs{\<q_1 q_2\> q} \>}
                        \hspace{-10mm}    p' \cdot E_{\bs{\sigma_1 q_1}}(n-1) \cdot E_{\bs{\sigma_2 q_2}}(n-1) \cdot E_{\bs{\<q_1 q_2\> q}}(n-1) \right. \\
   & \qquad \qquad \qquad \left. + \sum_{\bs{a q} \btran{p'} \bs{\sigma' q}} p' \cdot E_{\bs{\sigma' q}}(n-1) \right) \\
   & = \term{a}{q} \cdot E_{\bs{a q}}(n) \,,
 \end{align*}
 where the first and the last equality are by the definition of~$\Tim{}$ and the semantics of pSJSs,
  the first inequality and the second equality are trivial,
  the second inequality is by the induction hypothesis,
  and the third equality is by the definition of $\barGamma$.

 For the final statement, observe that
  \[
   \term{\bs{\sigma q}}{} = \lim_{n \to \infty} \P{\Tim{\bs{\sigma q}} \le n \mid \Run({\bs{\sigma q}})}
                         \ge \lim_{n \to \infty} \P{\Tim{\sigma} \le n \mid \tRun{\sigma}{q}}  = 1 \,.
  \]
\qed
\end{proof}

\subsection{Proof of Proposition~\ref{prop:bp-critical}}

Here is a restatement of Proposition~\ref{prop:bp-critical}.

\begin{qproposition}{\ref{prop:bp-critical}}
 \stmtpropbpcritical
\end{qproposition}

\begin{proof}
 Clearly, statement~(1) implies statement~(2), because, by definition, \mbox{$\Wor{X_0} \ge \Tim{X_0}$}.
 Next, we prove that statement~(3) implies statement~(1).
 For each $X \in \Gamma$ and $i \in \N$, we define a random variable~$\zs{i}_X$ over $\Run(X_0)$ by setting
  $\zs{i}_X(w) := |\Front(w(i))|_X$;
  i.e., $\zs{i}_X$ is the number of active $X$-processes at time~$i$.
 We assemble the $\zs{i}_X$ in a row vector~$\zs{i}$.
 Note that $|\Front(w(i))| = \norm{\zs{i}(w)}_1$.
 It is easy to see (by induction, see also \cite[p.~184]{AthreyaNey72}) that $\Ex{\zs{i}} = \es{X_0} \cdot A^i$,
  where by $\es{X_0} \in \N^\Gamma$ we mean the row vector whose only nonzero component is the $X_0$-component, which is~$1$.
 Consequently, we have
  \[
   \E \Wor{X_0} = \sum_{i=0}^\infty \Ex{|\Front(w(i))| \mid w \in \Run(X_0)} = \sum_{i=0}^\infty \norm{\Ex{\zs{i}}}_1
                = \norm{\es{X_0} \cdot A^*}_1 \,,
  \]
  where $A^* := \sum_{i=0}^\infty A^i$.
 It is known~\cite{book:HornJ} that the matrix series $A^*$ converges if and only if $\rho(A) < 1$.
 It follows that statement~(3) implies statement~(1).
 Further, it is known~\cite{book:HornJ} that if $\rho(A) < 1$, then $A^* = (I - A)^{-1}$, so then
  $\E \Wor{X_0} = \norm{\es{X_0} \cdot (I - A)^{-1}}_1$, which equals the $X_0$-component of $(I - A)^{-1} \cdot \vone$.

 It remains to show that statement~(2) implies statement~(3).
 For this part, we rely on Perron-Frobenius theory.
 Let $\rho(A) \ge 1$.
% (It follows from~\cite{EYstacs05Extended} that $\rho(A) \ge 1$, together with $\term{X_0}{} = 1$, implies $\rho(A) = 1$,
%  but we will not need that here.)
 Call a matrix~$B \in \R^{n \times n}$ \emph{strongly connected}, if for all $1 \le i,j \le n$ there is $k > 0$ such that $(B^k)_{i j} \ne 0$.
 Since $A$ is nonnegative, Corollary 2.1.6 of~\cite{book:BermanP} asserts that
  there exists a strongly connected principal submatrix $A'$ of~$A$ such that $\rho(A') \ge 1$;
 i.e., there is $\Gamma' \subseteq \Gamma$ such that the matrix $A' \in \R^{\Gamma' \times \Gamma'}$
  obtained from~$A$ by deleting all rows and columns not indexed with elements of~$\Gamma'$ is strongly connected.
 We will show that $\E \Tim{X} = \infty$ for all $X \in \Gamma'$.
 Since $(S, X_0)$ is reduced, this implies that $\E \Tim{X_0}$ is infinite.
 Therefore, to simplify the notation, we assume in the following w.l.o.g.\ that $A = A'$, i.e., $A$ is strongly connected.
 We will show $\E \Tim{X} = \infty$ for all $X \in \Gamma$.

 Define, for each $i \in \{0,1,2,3\}$, a function $\Ls{i}: (\R^\Gamma)^i \to \R^\Gamma$ as follows,
  where $\vy, \vz, \vw \in \R^\Gamma$ are column vectors (we use subscripts to refer to indices):
 \begin{align*}
  \Ls{0}_X & := \sum_{X \btran{p} \bot} p\,; & \Ls{2}(\vy,\vz)_X & := \sum_{X \btran{p} \<Y Z\>} p \cdot \vy_Y \cdot \vz_Z; \\
   \Ls{1}(\vy)_X & := \sum_{X \btran{p} Y} p \cdot \vy_Y;
    & \Ls{3}(\vy,\vz,\vw)_X & := \sum_{X \btran{p} \<Y Z \, W\>} p \cdot \vy_Y \cdot \vz_Z \cdot \vw_W.
 \end{align*}
 Notice that $\Ls{0}$ is a vector of constants, and $\Ls{1}, \Ls{2}, \Ls{3}$ are linear, bilinear, trilinear vector functions, respectively.
 We write $\Ls{2}(\vy,\cdot)$ and $\Ls{3}(\cdot,\vz,\vw)$ etc.\ to mean the matrices $U, V \in \R^{\Gamma \times \Gamma}$
  such that $U \cdot \vx = \Ls{2}(\vy,\vx)$ and $V \cdot \vx = \Ls{3}(\vx,\vz,\vw)$ etc.\ for all $\vx \in \R^\Gamma$.
 It is straightforward to verify that
  \[
   A  = \Ls{1}(\cdot) + \Ls{2}(\vone,\cdot) + \Ls{2}(\cdot,\vone) + \Ls{3}(\vone,\vone,\cdot) + \Ls{3}(\vone,\cdot,\vone) + \Ls{3}(\cdot,\vone,\vone) \,,
  \]
  where $\vone$ denotes the vector with all ones.
 Let $\vf : \R^\Gamma \to \R^\Gamma$ be the function with
  \[
   \vf(\vx) := \Ls{0} + \Ls{1}(\vx) + \Ls{2}(\vx,\vx) + \Ls{3}(\vx,\vx,\vx)\,.
  \]
 This ``generating function''~$\vf$ plays a central role in the branching process literature.
 Notice that $\vf$ characterises the branching process~$S$ up to the order of the children.
 Define $\qs{0} := \vzero$ (i.e., the vector with all zeros)
  and $\qs{i+1} := \vf(\qs{i})$ for all $i \in \N$.
 It is well-known~\cite{Harris63} (and straightforward to show by induction on~$i$) that $\P{\Tim{X} \le i} = \qs{i}_X$.
 Define $\rs{i} := \vone - \qs{i}$.
 Using either the definition of~$\rs{i}$ or the fact that $\rs{i}_X = \P{\Tim{X} > i}$) one can easily check
  \begin{align*}
   \rs{i+1} & = \Ls{1}(\rs{i}) + \Ls{2}(\vone,\rs{i}) + \Ls{2}(\rs{i},\vone-\rs{i}) \\
            & \qquad + \Ls{3}(\vone,\vone,\rs{i}) + \Ls{3}(\vone,\rs{i},\vone-\rs{i}) + \Ls{3}(\rs{i},\vone-\rs{i},\vone-\rs{i}) \\
            & = A \cdot \rs{i} - \Ls{2}(\rs{i},\rs{i}) - \Ls{3}(\vone,\rs{i},\rs{i}) - \Ls{3}(\rs{i},\vone,\rs{i}) \\
            & \qquad - \Ls{3}(\rs{i},\rs{i},\vone) + \Ls{3}(\rs{i},\rs{i},\rs{i}) \,.
  \end{align*}
 By defining
  \[
   B(\vx) := \Ls{2}(\vx,\cdot) + \Ls{3}(\vone,\vx,\cdot) + \Ls{3}(\vx,\vone,\cdot) + \Ls{3}(\vx,\cdot,\vone)
  \]
  we get
  \begin{equation}
   \rs{i+1}  = (A - B(\rs{i})) \cdot \rs{i} + \Ls{3}(\rs{i},\rs{i},\rs{i}) \,. \label{eq:rs}
  \end{equation}
 Note that $A - B(\vx)$ and $B(\vx)$ are nonnegative for $\vx \in [0,1]^\Gamma$
  and that $B(\varepsilon \cdot \vx) = \varepsilon \cdot B(\vx)$ for all $\varepsilon \in \R$.
% Hence, there is $b \in \R_+$ such that $\norm{B(\varepsilon \cdot \vone)}_\infty \le \varepsilon \cdot b$ for all $\varepsilon > 0$.

 In the following, for a vector $\vx$, we write $\vx_{\mathit{min}}$ and $\vx_{\mathit{max}}$ for the minimal and maximal entry of~$\vx$.
 It is easy to see that there is $s \in (0,1]$ such that for all $X,Y \in \Gamma$ we have $\P{\Tim{X} > n} \ge s \cdot \P{\Tim{Y} > n}$ for all $n \in \N$.
 (For instance, take for $s$ the probability to reach, starting in~$X$, a tree with~$Y$ in at most $|\Gamma|$ steps.
  This probability is positive, because $A$ is strongly connected.)
 It follows that $\rs{n}_{\mathit{min}} \ge s \cdot \rs{n}_{\mathit{max}}$.

 As $\rho(A) \ge 1$ and $A$ is strongly connected, Perron-Frobenius theory (see \cite{book:BermanP}) asserts that there is a vector $\vu \in (0,1]^\Gamma$,
  strictly positive in all components, such that $A \cdot \vu \ge \vu$.
 (For instance, one can take the dominant eigenvector of~$A$.)
\newcommand{\umin}{\vu_{\mathit{min}}}%
\newcommand{\umax}{\vu_{\mathit{max}}}%
 W.l.o.g.\ we can take $\umax = s$.
% We then have
%  \[
%   \norm{B(\varepsilon \cdot \vone) \cdot \vu}_\infty \le \norm{B(\varepsilon \cdot \vone)}_\infty \cdot s
%     \le \varepsilon \cdot \frac{b \cdot s}{\umin} \cdot \umin\,,
%  \]
%  hence, for $d := b \cdot s / \umin$, we have
 Choose $d \in \R_+$ such that
  \begin{equation}
    B(\vone) \cdot \vu \le d \cdot \vu\,. \label{eq:B-u}
  \end{equation}
 Define a sequence $(\varepsilon_n)_{n \in \N}$ by setting $\varepsilon_n := \rs{n}_{\mathit{max}}$.
 As $\rs{n}_{\mathit{min}} \ge s \cdot \varepsilon_n = \varepsilon_n \cdot  \umax$, we have
  \begin{equation}
   \varepsilon_n \cdot \vone \ge \rs{n} \ge \varepsilon_n \cdot \vu \,.
   \label{eq:rs-bounds}
  \end{equation}
 Observe that, since $A$ is strongly connected, we have
  $\P{\Tim{X} > n} > 0$ for all $X$ and all~$n$, and hence $\varepsilon_n \ge \varepsilon_{n+1} > 0$.
 If $(\varepsilon_n)_n$ does not converge to~$0$, then there is $c > 0$ and $X \in \Gamma$ such that $\P{\Tim{X} \ge n} \ge c$
  for all $n \in \N$, already implying that $\E \Tim{X} = \infty$.
 So we assume in the following that $\lim_{n \to \infty} \varepsilon_n = 0$.
\newcommand{\nbot}{n_\bot}
 In particular, there is $\nbot \in \N$ such that $\varepsilon_n d < 1$ for all $n \ge \nbot$.
 Now we show, for all $n \ge \nbot$ and all $i \in \N$, that
 \begin{equation}
  \rs{n+i} \ge (1 - \varepsilon_n d)^i \varepsilon_n \vu \,. \label{eq:bp-induction-claim}
 \end{equation}
 We proceed by induction on~$i$.
 The induction base ($i=0$) follows from~\eqref{eq:rs-bounds}.
 Let $i \ge 0$. We have:
 \begin{align*}
  \rs{n+i+1} & \ge (A - B(\rs{n+i})) \cdot \rs{n+i}                                                           && \text{(by~\eqref{eq:rs})} \\
             & \ge (A - B(\rs{n+i})) \cdot (1 - \varepsilon_n d)^i \varepsilon_n \cdot \vu                    && \text{(by induction hypothesis)} \\
             & \ge (1 - \varepsilon_n d)^i \varepsilon_n \cdot (\vu - B(\rs{n+i}) \cdot \vu)                  && \text{(as $A \vu \ge \vu$)} \\
             & \ge (1 - \varepsilon_n d)^i \varepsilon_n \cdot (\vu - B(\varepsilon_n \cdot \vone) \cdot \vu) && \text{(by~\eqref{eq:rs-bounds})} \\
             & \ge (1 - \varepsilon_n d)^i \varepsilon_n \cdot (\vu - \varepsilon_n d \cdot \vu)              && \text{(by~\eqref{eq:B-u})} \\
             & = (1 - \varepsilon_n d)^{i+1} \varepsilon_n \cdot \vu
 \end{align*}
 This proves~\eqref{eq:bp-induction-claim}.
 Now we have for all $n \ge \nbot$:
 \begin{align*}
   \sum_{i=0}^k \rs{n+i} & \ge \sum_{i=0}^k (1 - \varepsilon_n d)^i \varepsilon_n \vu                         && \text{(by~\eqref{eq:bp-induction-claim})} \\
                         &  =  \frac{1 - (1 - \varepsilon_n d)^{k+1}}{1 - (1 - \varepsilon_n d)} \varepsilon_n \vu \\
                         &  =  \frac{1 - (1 - \varepsilon_n d)^{k+1}}{d} \cdot \vu \,,
 \end{align*}
  so, for every $n \ge \nbot$ there exists some $k(n) \in \N$ such that $\sum_{i=n}^{k(n)} \rs{i} \ge \frac{1}{2d} \cdot \vu$.
 Hence, for any $X \in \Gamma$ we have
 \begin{align*}
  \E \Tim{X} & = \sum_{i=0}^\infty \P{\Tim{X} > i} = \sum_{i=0}^\infty \rs{i}_X \ge \sum_{i=\nbot}^\infty \rs{i}_X
                 = \sum_{i=\nbot}^{k(\nbot)} \rs{i}_X + \sum_{i=k(\nbot)+1}^{k(k(\nbot)+1)} \rs{i}_X + \cdots \\
             & \ge \frac{1}{2d} \vu_X + \frac{1}{2d} \vu_X + \cdots = \infty \,,
 \end{align*}
  because $\vu_X > 0$.
 This completes the proof.
\qed
\end{proof}

\subsection{Proof of Corollary~\ref{cor:bp-work-time}}

Here is a restatement of Corollary~\ref{cor:bp-work-time}.

\begin{qcorollary}{\ref{cor:bp-work-time}}
 \stmtcorbpworktime
\end{qcorollary}
\begin{proof}
 By Proposition~\ref{prop:bp-critical}, the expectations $\E \Wor{X_0}$ and $\E \Tim{X_0}$ are both finite or both infinite.
 To distinguish between those cases, compute in polynomial time the matrix~$A$.
 As $A$ is nonnegative, we have $\rho(A) \ge 1$ if and only if there is a nonnegative vector~$\vx$ which is nonzero in at least one component
  and $A \vx \ge \vx$ holds (i.e., ``${\ge}$'' holds in all components), see~\cite{book:BermanP}.
 Therefore, $\rho(A) \ge 1$ if and only if the linear programming (LP) problem ``$A \vx \ge \vx \ge \vzero$ and $\norm{\vx}_1 = 1$'' is feasible.
 This can be decided in~\PTime.
 We remark that a similar method was used in~\cite{EYstacs05Extended}.
\qed
\end{proof}

\subsection{Proof of Theorem~\ref{thm:work-time}}

Here is a restatement of Theorem~\ref{thm:work-time}.

\begin{qtheorem}{\ref{thm:work-time}}
 \stmtthmworktime
\end{qtheorem}
\begin{proof}
 If $S$ is not normalised, we normalise it in polynomial time (Section~\ref{sub:termination-probabilities}).
 This does not change the finiteness of $\E \Wor{a}$ or~$\E \Tim{a}$.
 Then we compute in polynomial time (Theorem~\ref{thm:termination-probabilities}) the set $Q' := \{ q \in Q \mid \term{a}{q} > 0\}$.
 The values $\term{a}{q}$ for $q \in Q'$ can be efficiently expressed (Theorem~\ref{thm:termination-probabilities}).
 Therefore, we can also efficiently express $\term{a}{} = \sum_{q \in Q'} \term{a}{q'}$
  and decide in \PSPACE\ if $\term{a}{} = 1$ or $\term{a}{} < 1$.
 If $\term{a}{} < 1$, then nonterminating runs have a positive probability and hence $\E \Wor{a} = \E \Tim{a} = \infty$.
 Otherwise, we have, with Proposition~\ref{prop:transformation},
\newcommand{\tilW}{\widetilde{W}}%
\newcommand{\tilT}{\widetilde{T}}%
 \begin{align}
  \E \Wor{a} & = \sum_{q \in Q'} \term{a}{q} \cdot \Ex{\Wor{a} \mid \tRun{a}{q}}
    = \sum_{q \in Q'} \term{a}{q} \cdot \E \Wor{\bs{a q}} =: \tilW  \label{eq:tilW} \\
  \text{and} \quad \E \Tim{a} & = \sum_{q \in Q'} \term{a}{q} \cdot \Ex{\Tim{a} \mid \tRun{a}{q}}
   \ge \sum_{q \in Q'} \term{a}{q} \cdot \E \Tim{\bs{a q}} =: \tilT \,. \label{eq:tilT}
 \end{align}
 If $\E \Wor{a}$ is finite, then $\E \Tim{a}$ is finite, because $\Wor{a} \ge \Tim{a}$.
 If $\E \Tim{a}$ is finite, then $\tilT$ is finite by~\eqref{eq:tilT}, hence $\tilW$ is finite by Corollary~\ref{cor:bp-work-time},
  hence $\E \Wor{a}$ is finite by~\eqref{eq:tilW}.
 Therefore, $\E \Wor{a}$ is finite if and only if $\E \Tim{a}$ is finite.

 In order to decide if $\E \Wor{a}$ is finite, by~\eqref{eq:tilW} it suffices to decide if $\E \Wor{\bs{a q}}$ is finite for all $q \in Q'$.
 We cannot use Corollary~\ref{cor:bp-work-time} directly,
  because the coefficients of the branching process from Proposition~\ref{prop:transformation} are not explicitly given.
 However, by Theorem~\ref{thm:termination-probabilities} they are efficiently expressible and, therefore, so is the matrix~$A$
  from Proposition~\ref{prop:bp-critical}.
 So we can decide in \PSPACE\ whether $\E \Wor{\bs{a q}}$ is finite by deciding whether the formula
  ``$\exists \vx : A \vx \ge \vx \ge \vzero$ and $\norm{\vx}_1 = 1$'' in~$\ExThR$ is true (cf.\ the proof of Corollary~\ref{cor:bp-work-time}).

 If $\E \Wor{a}$ is finite, we wish to efficiently express it.
 We use again $\tilW$ from~\eqref{eq:tilW},
  so it suffices to show that $\E \Wor{\bs{a q}}$ is efficiently expressible for all $q \in Q'$.
 For that, notice again that the matrix~$A$ from Proposition~\ref{prop:bp-critical} is efficiently expressible,
  and consider the formula ``$(I - A) \vx = \vone$'' in~$\ExThR$.
 Existentially quantify all variables in~$\vx$ except the $\bs{a q}$-component.
 By Proposition~\ref{prop:bp-critical}, the resulting formula efficiently expresses $\E \Wor{\bs{a q}}$.

 It remains to show \PosSLP-hardness for pPDSs.
 We draw from a reduction in~\cite{EYstacs05Extended}, where it is shown that deciding if a pPDS terminates with probability~$1$ is \PosSLP-hard.
 As a gadget for that,
  Etessami and Yannakakis~\cite{EYstacs05Extended} compute, given a \PosSLP\ instance, a pPDS $S$ with the following properties.
 (More precisely, they construct an equivalent recursive Markov chain.)
 The starting configuration is $qa$, and after having left the initial configuration,
  $S$ reaches a configuration of the form $q \alpha$ with $\alpha \in \Gamma^*$ again with probability~$1$.
 At that time, the configuration is~$q$ with some probability~$p$, and $qaa$ with probability $1-p$.
 Moreover, the time (and hence work) needed to reach either of those configurations
  is essentially bounded by the size of the given \PosSLP\ instance, so it is finite.
 Furthermore, the given \PosSLP\ instance is a ``yes instance'' if and only if $p > \frac12$.
 It is easy to see that $\E \Wor{qa}$ is finite in~$S$ if and only if $\E \Wor{X}$ is finite in the branching process~$\barS$ that consists only
  of the transitions $X \btran{p} \bot$ and $X \btran{1-p} \<X X\>$.
 The $1 \times 1$-matrix~$A$ for~$\barS$ from Proposition~\ref{prop:bp-critical} consists of a single entry $2 \cdot (1-p)$.
 Consequently, $\E \Wor{qa}$ is finite in~$S$ if and only if $2 \cdot (1-p) < 1$, which is equivalent to $p > \frac12$.
 This completes the reduction.
 We remark that the reduction does not show that deciding if $\E \Wor{}$ is finite is \PosSLP-hard for branching processes,
  because $S$ has more control states than just~$q$.
 (Recall that the problem for branching processes is in~\PTime\ by Corollary~\ref{cor:bp-work-time}.)
\qed
\end{proof}

\subsection{Proof of Proposition~\ref{prop:seneta}}

Here is a restatement of Proposition~\ref{prop:seneta}.

\begin{qproposition}{\ref{prop:seneta}}
 \stmtpropseneta
\end{qproposition}
\begin{proof}
 By Proposition~\ref{prop:bp-critical}, we have $\E \Wor{X} = 1 / (1-2p)$.
 It is shown in~\cite{Seneta68} that
  $\lim_{p \to \frac12} \frac{\E \Tim{X}}{-2 \ln(1-2p)} = 1$.
 Consequently, we have $\lim_{p \to \frac12} \Ex{\Wor{X}} / \Ex{\Tim{X}} = \infty$.
\qed
\end{proof}